# Mid Frequency Aperture Array Architectural Design Document

Authors: A.W. Gunst, A.J. Faulkner, S.J.  Wijnholds, R. Jongerius, S. Torchinsky, W.A. van Cappellen

# Context

The Square Kilometre Array (SKA) is the next generation radio telescope. Aperture Arrays (AA) are considered for SKA-2 for frequencies up to 1.4 GHz (SKA-1 uses AAs up to 350 MHz). This document presents design considerations of this Mid-Frequency Aperture Array (MFAA) element and possible system architectures complying with the SKA-2 system requirements, combining high sensitivity with a superb survey speed.
The architectural analyses has been submitted to the System Requirements Review of the MFAA element.

# TABLE OF CONTENTS

# LIST OF FIGURES

# LIST OF TABLES

# LIST OF ABBREVIATIONS

| | |
|---|---|
| AAMID | Aperture Array MID frequency system or consortium |
| ADC | Analog to Digital Converter |
| AD-n | nth document in the list of Applicable Documents |
| CPF | Central Processing Facilities |
| CSP | Central Signal Processor |
| DN | Data Network |
| EMBRACE | Electronic Multi Beam Radio Astronomy ConcEpt |
| EMI | Electro Magnetic Interference |
| ESD | Electro Static Discharge |
| Flops | Floating Point Operation per second |
| FoV | Field of View |
| ICD | Interface Control Document |
| IT | Information Technology |
| INFRA-SA | Infrastructure Africa |
| KAT | Karoo Array Telescope |
| LFAA | Low Frequency Aperture Array |
| LNA | Low Noise Amplifier |
| LOFAR | Low Frequency Aperture Array |
| MAC | Multiply and Accumulate |
| MCCS | Monitor, Control and Calibration System |
| MeerKAT | “More” KAT |
| MFAA | MID Frequency Aperture Array |
| MWA | Murchison Widefield Array |
| OSI | Open Systems Interconnection |
| PBS | Product Breakdown Structure |
| RD-N | nth document in the list of Reference Documents |
| RF | Radio Frequency |
| RFI | Radio Frequency Interference |
| RPF | Remote Processing Facilities |
| SA | South-Africa |
| SNR | Signal to Noise Ratio |
| SDP | Science Data Processing |
| SKA | Square Kilometre Array |
| SKAO | SKA Office |
| SPM | Station Processing Modules |
| SRR | System Requirements Review |
| TBD | To Be Done |
| TECU | Total Electron Content Unit |
| TM | Telescope Management |
| TPM | Tile Processor module |
| TRL | Technology Readiness Level |
| WAN | Wide Area Network |

# 1 Introduction

## 1.1 Purpose of the document

The objective of this document is to present design considerations of the Mid-Frequency Aperture Array element and possible system architectures that meet the MFAA system requirements. In this document the requirements of [RD-1] are mapped onto functions to show that feasible designs can be built. Since the time to construction is far away and the available technology at that time is only projected, several architectural options are presented.

## 1.2 Scope of the document

The scope of this document is the MFAA system. One of the sections broadens the scope to the full AAMID system to show the feasibility of the AAMID telescope as a whole.

# 2 References

## 2.1 Applicable documents

The following documents are applicable to the extent stated herein. In the event of conflict between the contents of the applicable documents and this document, **the applicable documents** shall take precedence.

| Id | Title | Code | Issue |
|---|---|---|---|
| AD-1 | SKA Request for proposal Proposals | SKA-TEL.-SKO-0000020 | 01 |
| AD-2 | Statement of work for the study, prototyping and design on an SKA element | SKA-TEL.-SKO-0000021 | 01 |
| AD-3 | Statement of Work for the Study, Prototyping and Preliminary Design of an SKA Advanced Instrumentation Programme Technology | SKA-TEL.-SKO-0000022 | 01 |
| AD-4 | SKA Pre construction Top Level WBS | SKA-TEL.-SKO-0000023 | 01 |
| AD-5 | SKA-1 System Baseline design | SKA-TEL.-SKO-0000002 | 01 |
| AD-6 | The Square Kilometre Array Design Reference Mission: SKA Phase 1 | SKA-TEL.-SKO-0000002 | 03 |
| AD-7 | SKA System Engineering Management Plan | SKA-TEL.-SKO-0000024 | 1 |
| AD-8 | The Square Kilometre Array Intellectual Property Policy | SKA-GOV-0000001 | 1.3 Draft |
| AD-9 | Draft Consortium Agreement | PD/SKA.26-4 | Draft |
| AD-10 | Document Requirements Description | SKA-TEL.-SKO-0000029 | 1 |
| AD-11 | SKA Document Management Plan | SKA-TEL.-SKO-0000026 | 2 |
| AD-12 | SKA Product Assurance and Safety Plan | SKA-TEL.-SKO-0000027 | 1 |
| AD-13 | Change Management Procedure | SKA-TEL.-SKO-0000028 | 1 |
| AD-14 | SKA Interface Management Plan | SKA-TEL.-SKO-0000025 | 1 |
| AD-15 | SKA Phase 1 System (Level 1) requirements specification | SKA-TEL.-SKO-0000008 | 3 |

## 2.2 Reference documents

The following documents are referenced in this document. In the event of conflict between the contents of the referenced documents and this document, **this document** shall take precedence.

| Id | Title | Code | Issue |
|---|---|---|---|
| RD-1 | SKA-AAMID System Requirements | SKA-TEL-MFAA-0200005 | 1.0 |
| RD-2 | MFAA Requirements | SKA-TEL-MFAA-0200008 | 1.0 |
| RD-3 | S.J. Wijnholds, *"Fundamental Accuracy Limits of Ionospheric Characterization by Radio Interferometers,"* URSI Atlantic Radio Science Conference (URSI AT-RASC), 16-24 May 2015. | DOI: 10.1109/URSI-AT-RASC.2015.7303209 | |
| RD-4 | SKA-AAMID System Requirements | SKA-TEL-MFAA-0200005 | 1.0 |
| RD-5 | EMBRACE: A Multi-Beam 20,000-Element Radio Astronomical Phased Array Antenna Demonstrator | IEEE Transactions on Antennas and Propagation, volume: 59, issue: 6 | |
| RD-6 | Antenna Noise Temperature Calculation | Memo 95 | |
| RD-7 | APERTIF: Phased Array Feeds for the Westerbork Synthesis Radio Telescope | IEEE International Symposium on Phased Array Systems and Technology, October 2010 | |
| RD-8 | AA-Mid Concept Descriptions | WP2-015.020.010-TN-001 | D |
| RD-9 | Aperture Array Integrated Receiver | French funding proposal 2011 | |
| RD-10 | An end-to-end computing model for the Square Kilometre Array | IEEE Computer, vol. 47, no. 9, pp. 48–54, Sept 2014 | |
| RD-11 | PDR.05 parametric models of SDP compute requirements | SKA-TEL-SDP-0000040 | |
| RD-12 | MFAA Technology Description | SKA-TEL-MFAA-0200006 | B |
| RD-13 | MFAA Cost and Power Estimate | SKA-TEL-MFAA-0100006-MGT-CRE-B | B |
| | | | |
| | | | |
| | | | |

# 3 Introduction

The Mid Frequency Aperture Array (MFAA) within the scope of the AAMID telescope is depicted in Figure 1. It consists basically of all equipment in the field and processing facility to produce the final station voltage beams. The prime function is to receive Electromagnetic (EM) waves from the sky and accordingly select a piece of the solid angle on the sky in order to reduce the data to manageable levels. This results in station voltage beams. The concept of the "station" is used for grouping multiple antennas together. The station voltage beams of all stations are correlated together for the imaging applications resulting in visibilities. An important property of the correlator is to reduce the data rate significantly by time averaging. Since the time resolution of the pulsar applications need to be much more fine, a Tied Array Beamformer (TAB) in the central signal processor takes care of a weighted addition of a sub-set of the stations (primary the core). Therefore, the output product of the central signal processor consists of visibilities and/or beams. These are fed to the science data processor, which performs post processing, including array calibration, source subtraction and image cleaning amongst others. Also the de-dispersion functions are typically assigned to the science data processor. Although currently the functions of the central signal processor and science data processor are strictly separated this might change in the future. Due to the inherent flexibility of the MFAA some functions might be re-ordered for some specific applications. The AAMID telescope is controlled by a telescope management system. Infrastructure provides the telescope with power, roads, networks, clock and timing.

The rest of the document is primarily focused on the MFAA.

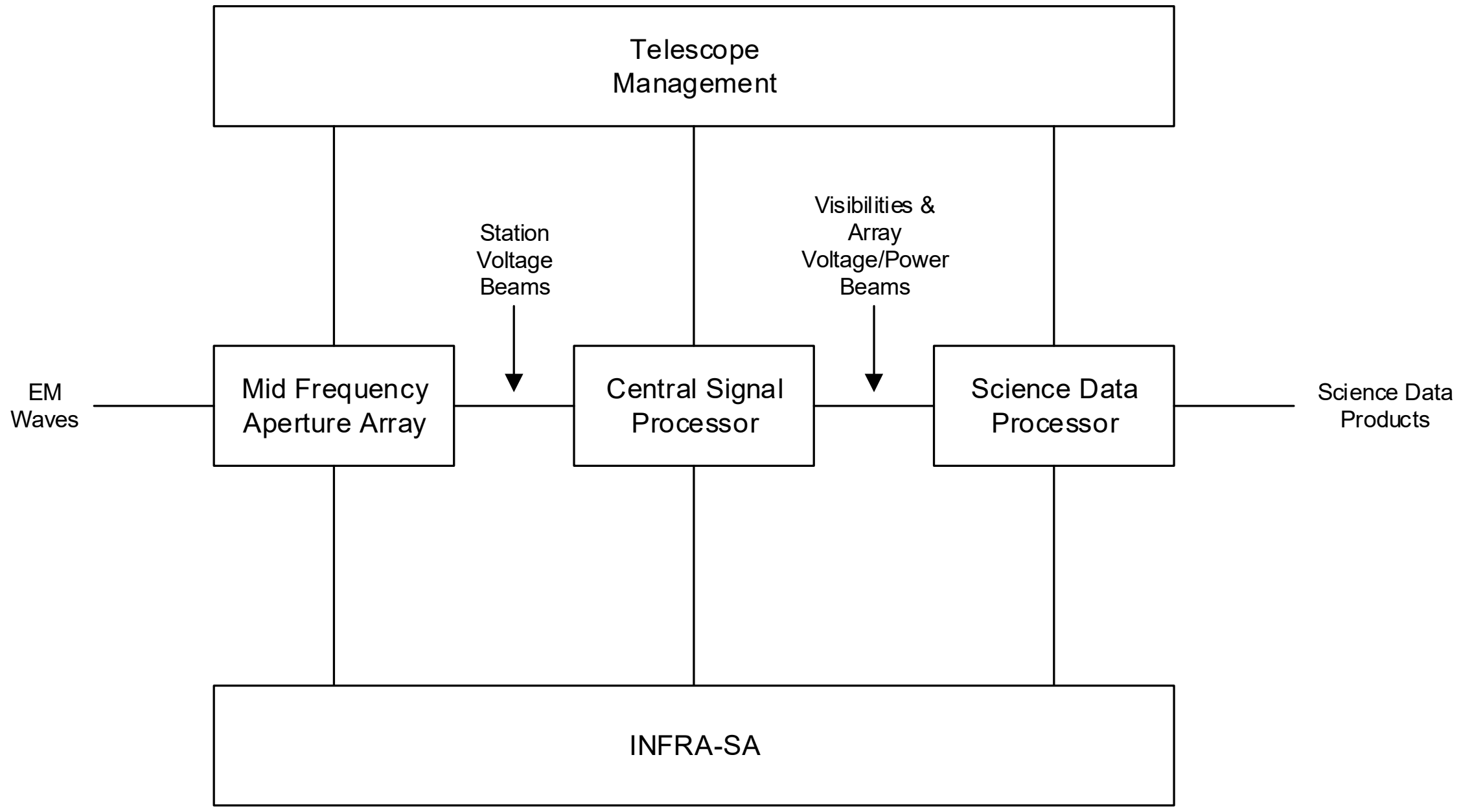

**Figure 1: MFAA in the AAMID Telescope**

# 4 AAMID Overall System Description

SKA-AAMID is a complete mid-frequency radio telescope. It is a very high performance "IT" based telescope in that there are no moving parts and the configuration for observations is controlled by software. With careful design of the architecture of the overall telescope alternative processing scenarios can be developed that would be designed to improve the performance substantially. A simplified outline of the full SKA-AAMID is shown in Figure 2. MFAA is the collector technology shown on the left, this is the principle subject of the architecture reported here. However, MFAA is a part of the overall telescope where useful trade-offs in processing and algorithms can be applied.

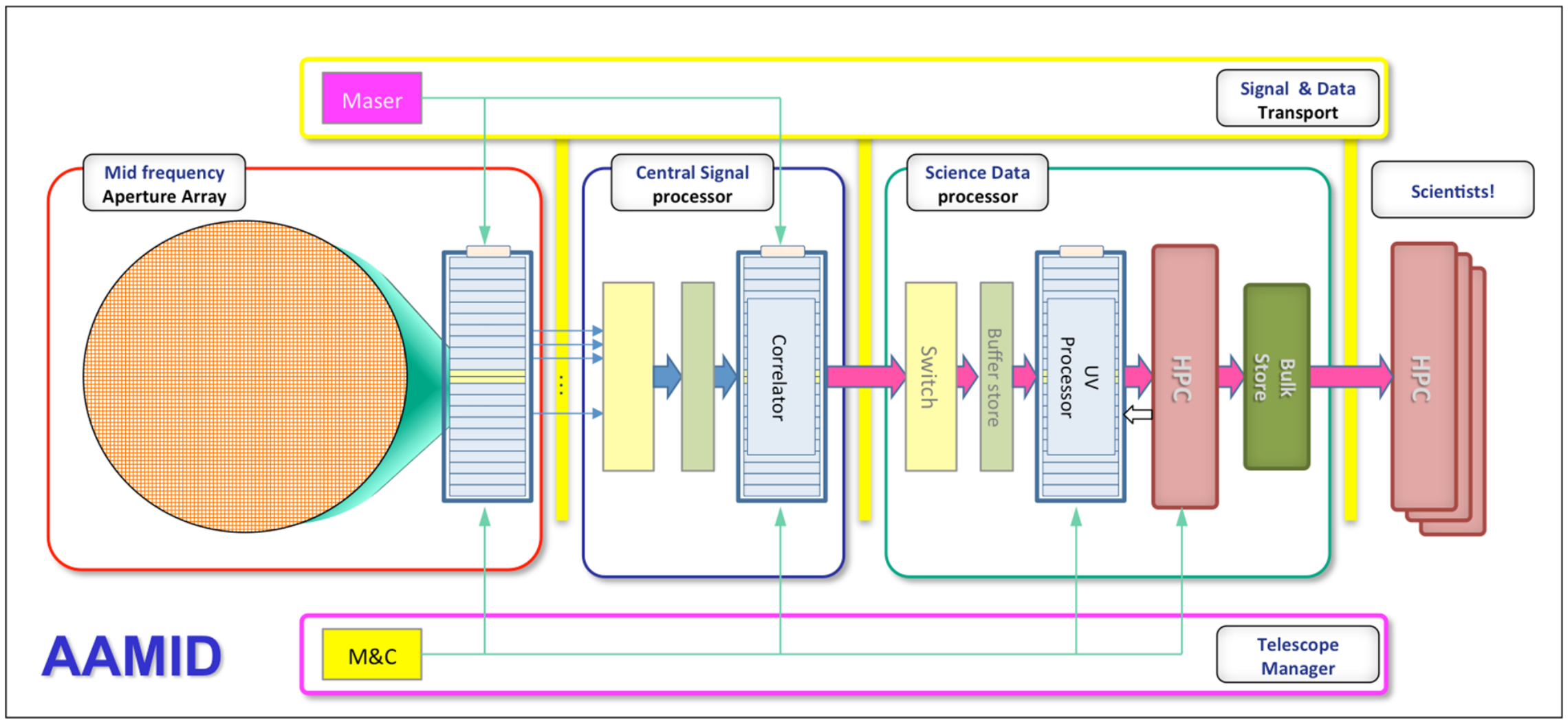

**Figure 2: SKA-AAMID outline block diagram**

The structure of AAMID is following a conventional radio telescope approach whereby the collector, MFAA, receives signals from the sky and forms beams on the sky. MFAA during observation will be formed into many, ~250 stations, that can each form beams for correlation and subsequent imaging or time domain processing. In the case of MFAA these can be very many beams for high survey speed and flexibility of operation.

The beams formed by MFAA are sent to a correlator, which can be split into as many partitions as there are beams from each of the MFAA stations. The correlator performs conventional correlation and integration into visibilities, which are then passed onto the Science Data Processor to form into image cubes, or time domain information that is stored for subsequent science processing dependent upon the experiment.

As can be seen this is all information processing either in specialist digital signal processing or high performance computing centres. There are no moving parts. The subsystems of AAMID are on the primary development path of the IT industry, so performance and upgrades will be available over time.

The performance capability of AAMID is very high, with a total beams*bandwidth product of at least 50GHz and maybe as high as 250GHz. This is the same as having 50 off 1GHz beams on the sky, or 100 off 500MHz beams – this is a lot of data!

The advantages of the IT telescope come into their own when considering flexibility of configuration and use. Multiple beams of arbitrary bandwidth can be formed to tailor a survey to the science experiment's requirements; different experiments can take place concurrently using their own beam(s); it is reasonable to dynamically reconfigure the array – size of stations etc. – for specific observations; the array can be repointed very quickly with no slewing delays as with mechanical pointing.

Calibration is a key requirement for all radio telescopes. The MFAA can calibrate the effective receiving "surface" with exquisite precision as required. This is due to the ability of the digital systems to adjust each signal path to compensate for any analogue errors as a function of frequency.

Upgrades of this telescope can take place using progressively better IT equipment. This will all be housed in "normal" racks in processing facilities. Maybe of more benefit is that it will be possible to change the processing architecture to reduce the post processing required: the concept of beamforming to correlation to gridding etc. is based on the use of dishes – which are inherent mechanical beamformers. For example, correlation at antenna/tile level followed by beamforming may have significant benefits. With a flexibly configured IT architecture of the whole telescope then many alternatives can be proposed and tried. This can be the basis of making advances in radio astronomy techniques.

# 5 AAMID Requirements

In this section the driving science requirements are identified and presented. Next, a number of technical requirements are presented more thoroughly.

## 5.1 Driving Requirements

In this section the envelope of the MFAA requirements from [RD-1] are derived by taking the maximum value for each of the science requirements as discussed in [RD-1]. These are input to the design phase, wherein several trade-offs are made in order to achieve the requirements. Figure 3 shows the survey speed requirement as a function of frequency. Therein can be seen that the required survey speed is flat.

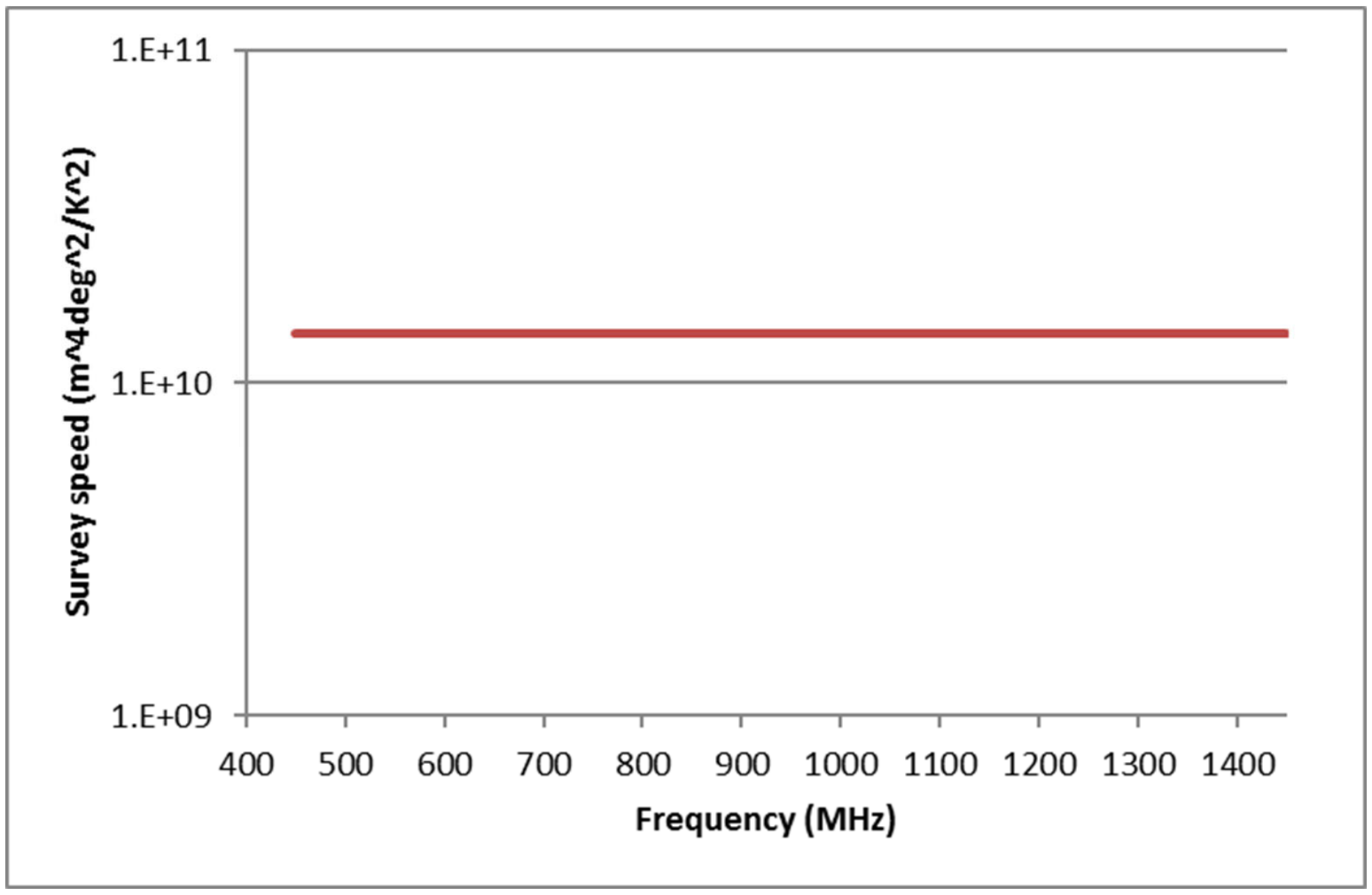

**Figure 3: Survey speed requirement as a function of frequency**

The envelope A/T requirement is depicted in Figure 4, which is also flat.

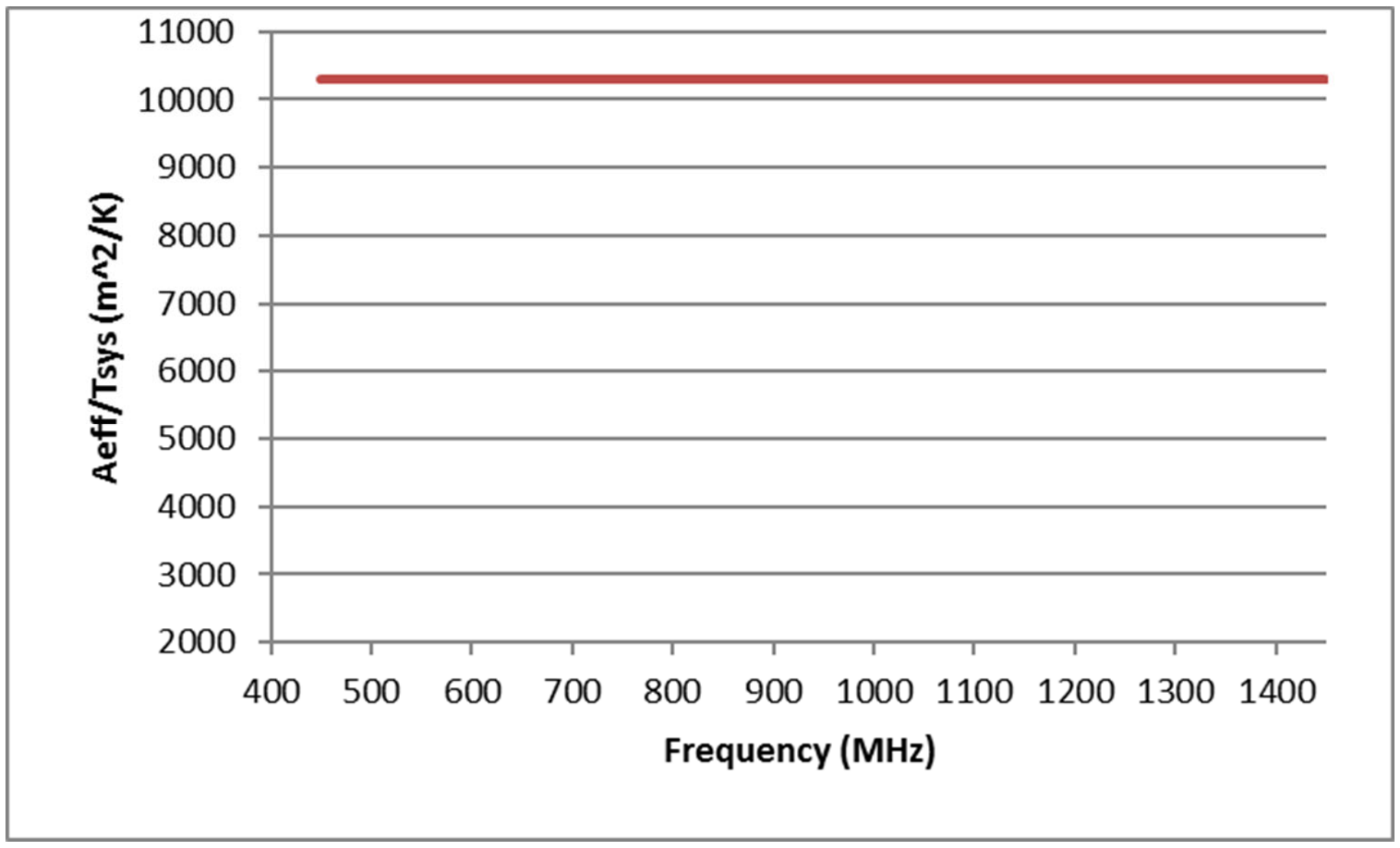

**Figure 4: A/T requirements as a function of frequency**

In Figure 5 the FoV requirement is depicted. The curve with 250 MHz as parameter is the requirement from the transients. The other curves show the reduced FoV in case the instantaneous bandwidth is larger. A $f^{-2}$ profile is assumed, wherein f is the frequency.

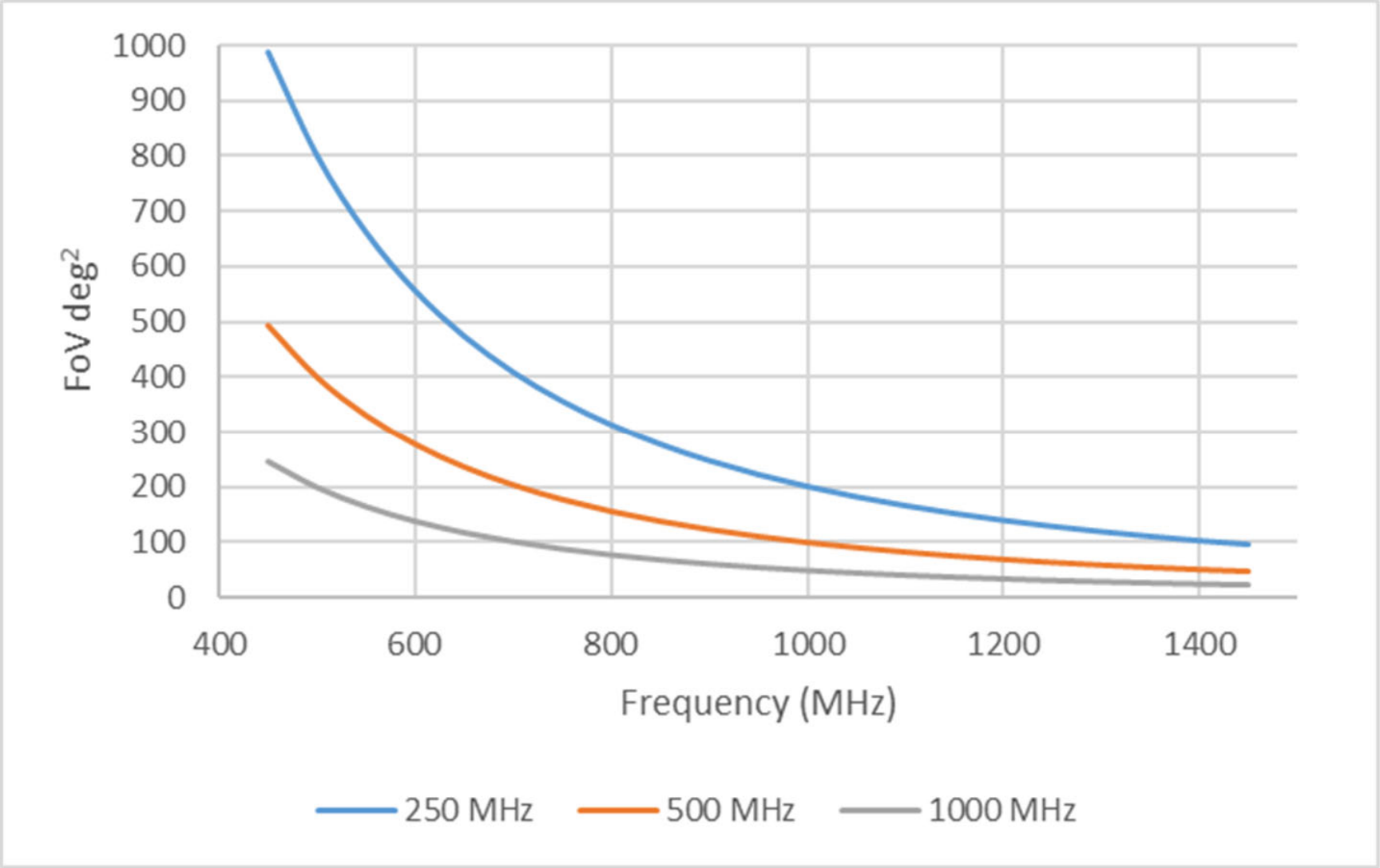

**Figure 5: Solid angle requirement as a function of frequency. The instantaneous bandwidth is used as parameter.**

Based on these AAMID requirements, realistic technical requirements have been set. These are listed in **Table 1** and more thoroughly discussed in the next sub-sections.

| Parameter | Essential | Desirable | Unit |
|---|---|---|---|
| Lowest frequency | 450 | 400 | MHz |
| Highest frequency | 1450 | 1500 | MHz |
| Instantaneous bandwidth | 1000 | 1100 | MHz |
| Optical FoV @ 1 GHz | > 200 | > 250 | sq. deg |
| Processed FoV @ 1 GHz (500 MHz bandwidth) | > 100 | | sq. deg |
| Bandwidth beam product @ 1 GHz | 500 * 100 | | MHz sq. deg |
| Trade-off flexibility | bandwidth for FoV | | |
| Polarisation | 2 | | |
| A/T target | > 10,000 | > 15,000 | m^2/K |
| Survey speed | > 1.4e10 | | sq. deg m^4/K^2 |
| Buffer capability (> 250 MHz bandwidth) | > 120 | | seconds |
| Minimal baseline | < 20 | < 1 | m |
| Maximal baseline | > 309 | > 515 | km |
| Scan angle from zenith | ±45 | ±60 | deg |

**Table 1: Technical requirements**

## 5.2 Bandwidth and Field of View

The required frequency range for AAMID is set from 450 - 1450 MHz, with a best effort to lower frequencies towards 400 MHz and higher frequencies towards 1500 MHz. This frequency range needs to be covered by the full MFAA signal processing chain. At most an instantaneous bandwidth of in total 1 GHz needs to be achieved to cover the band continuously.

The optical FoV is defined as the maximum FoV which can be "seen" by the telescope instantaneously. This is limited by the maximum beamwidth that can be formed in the analog domain. The optical FoV has to be larger than or equal to the processed FoV. The latter is the FoV which the system is able to process into "station beams". There are two reasons to distinguish between them:

1. Station beams can be further apart, with gaps, and do not necessary have to create a processed FoV contiguously.
2. The processing cost of processing all the optical FoV might be too large in the first years of operation. Later on this capacity can be extended to process more of the optical FoV.

To operate as a versatile telescope, flexibility in the system is required to trade bandwidth for processed FoV. The limit of the beam-band product is set by an overarching requirement specifying the combination of both the beam-width and bandwidth. For this a FoV profile over frequency is assumed which follows an $f^{-2}$ relation as function of frequency $f$. Specifying the FoV at a frequency of 1 GHz results in a required beam-band product of 100 * 500 deg$^2$ MHz. This means an FoV profile with 100 deg$^2$ at 1 GHz can be achieved for an instantaneous bandwidth of 500 MHz. Henceforth a FoV profile of 200 deg$^2$ could be achieved over half the bandwidth (250 MHz).

Although the natural FoV profile follows an $f^{-2}$ relation as a function of frequency, other FoV profiles can be created as well, due to the flexibility of the MFAA system. Figure 6 shows various profiles which can be created from the beam-band product of 100 * 500 deg$^2$ MHz given the same instantaneous bandwidth of 500 MHz. For example, a science case requiring a $f^{-2}$ FoV profile (blue curve) with 100 deg$^2$ at 1 GHz would need the same beam-bandwidth product as a science case requiring a 100 deg$^2$ of FoV, flat over the band (green curve).

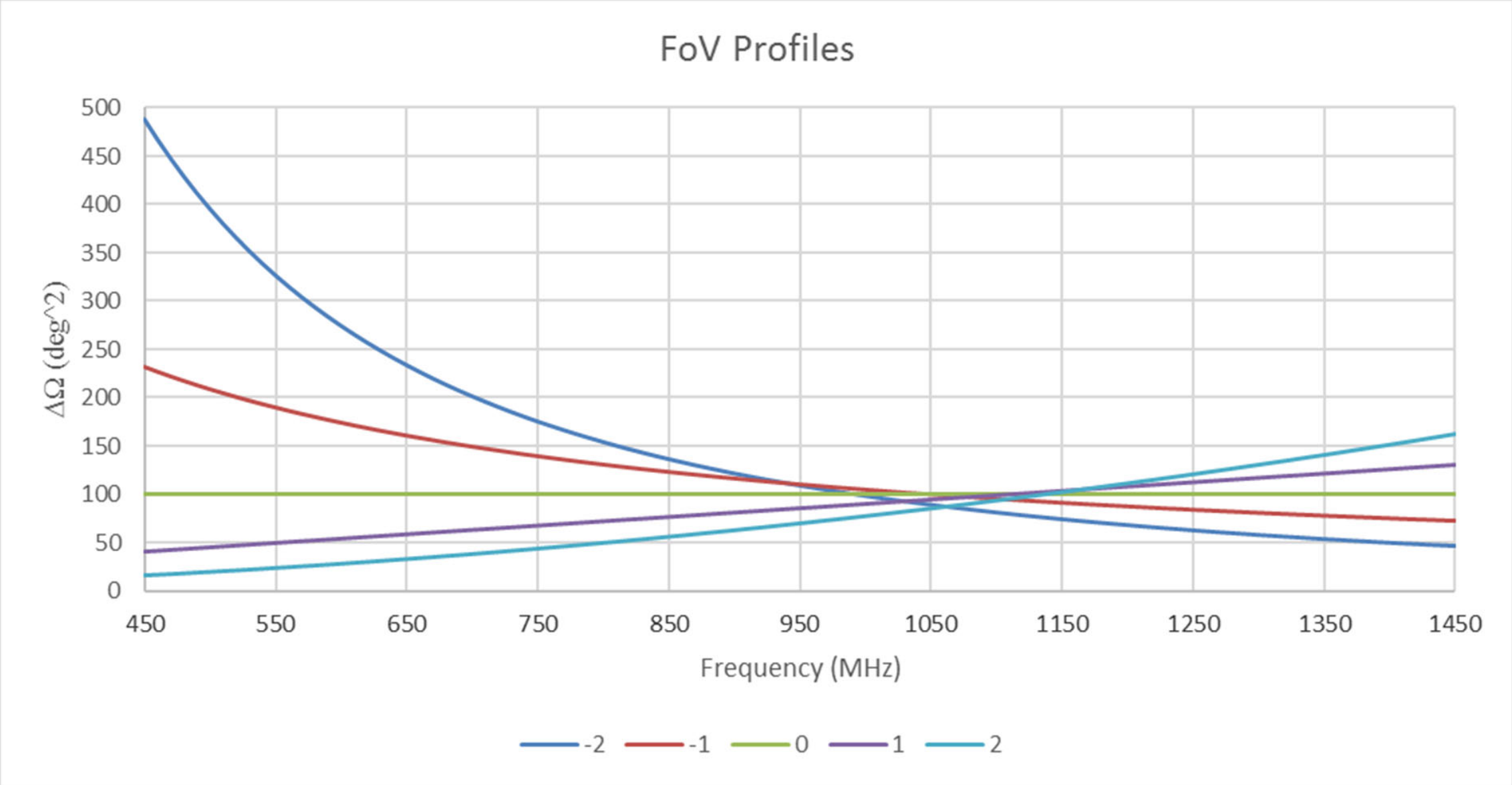

**Figure 6: Possible FoV profiles for an instantaneous bandwidth of 500 MHz with the power, *x*, of the frequency as parameter to yield a frequency profile of $f^x$**

## 5.3 Polarisation

Two polarisations are specified. These can be used to determine the polarisation properties of objects, but also to enhance the sensitivity by a factor $\sqrt{2}$ for unpolarised sources.

## 5.4 Sensitivity and Survey Speed

In general, the sensitivity requirement is driven by science cases, which do not need to cover a large piece of the sky [RD-1], while the survey speed requirements are driven typically by surveys. Of course the surveys also benefit from sensitivity, since the survey speed increases with the square of the sensitivity. Table 1 specifies a requirement for both within the band.

## 5.5 Buffer Capability

The purpose of a buffer is to store raw antenna data. A buffer capability of at least 120 seconds is required for a bandwidth of 250 MHz to cover dispersion measures over a large part of the bandwidth. The buffer time is also sufficient to accommodate for latency in the signal chain and pipelines to trap transients. The raw antenna signals buffered can be composed from multiple antennas to save on total buffer memory.

The buffer memory should be flexibly assigned to give much longer storage times over narrower bandwidths and indeed it offers more FoV than is normally processed, since the raw combined antenna signals are stored before beamforming.

## 5.6 Baseline Length

The baseline length requirements come from several science cases. Some science cases need long baselines, while others need extremely short baselines. A major advantage of a flexible aperture array instrument is its ability to re-configure the whole instrument to be tailored to a specific science case. The extremely short baselines can be realized by re-configuring the system in much smaller logical stations. These experiments typically only need the core area. The limit of the amount of smaller stations to correlate is obviously limited by the processing further up in the signal chain (correlator and science data processing).

## 5.7 Scan Angle

A scan angle of -60 and + 60 degrees from zenith is desirable to ensure a sky coverage of 30,000 $deg^2$ within 24 hours is visible for the telescope. The required scan angle is ±45 degrees, resulting in a sky coverage of 25,000 deg2. The sensitivity degradation with scan angle particularly up to 60 degrees will depend on the array design. The embedded element pattern describes the degradation of the sensitivity over scan angle, irrespective of the array architecture. This includes all effects, including mutual coupling.

# 6 MFAA Architecture

In this section a generic MFAA architecture is discussed. Various options to implement this architecture will be further discussed in Section 9.

## 6.1 External Interfaces

The external interfaces for the MFAA are shown in Figure 7 and further discussed in the following subsections.

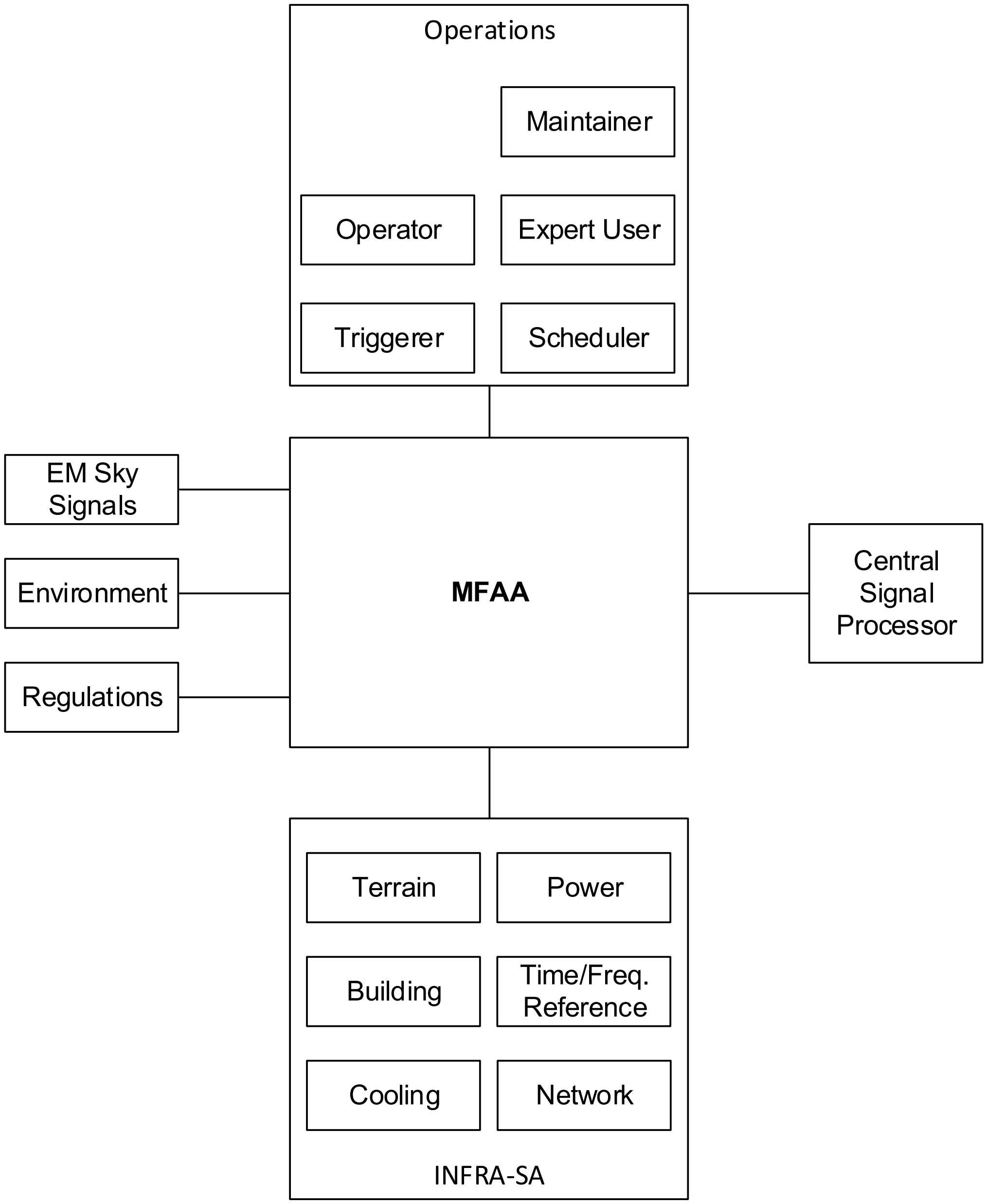

**Figure 7: External interfaces**

### 6.1.1 EM Sky Signals

This interface consists of the electro-magnetic signals from the sky. The word sky is used to discriminate these signals from human made signals such as Radio Frequency Interference (RFI). The RFI signals are part of the environment interface.

### 6.1.2 Environment

The MFAA system will be built in the outside and has to face the various conditions of the environment. This interface will specify that further in terms of environment. The environment encompasses the weather, RFI, the particular conditions of the site like the presence of sand, animals and any other external influence.

### 6.1.3 Regulations

The interface with regulations, covers all legal regulations which needs to be followed. This ranges from safety regulations to regulations concerning the emission of EM signals at the site.

### 6.1.4 Telescope Management to MFAA Interface

The telescope management system schedules MFAA resources and prepares the MFAA system for observations. It interfaces with the monitor & control system of MFAA, which takes care of the translation of telescope management functions into instructions for the underlying hardware. Typically instrument configurations are translated into hardware specific parameters, for example the translation of the source coordinates to weights and settings of the beamformers.

The interface is also used for monitoring the status of the underlying hardware and signal chain. Statistics can be retrieved from the MFAA monitoring & control system to enable deep inspections of the underlying data flows.

More specifically the interface can be grouped in the following underlying interfaces:

- Operator: this interface defines the requirements from an operator point of view. Typical high level control functions are available for the operator, as well as a higher level of monitoring and log information. Alarms are visible for the operator in case of malfunction and action is required.
- Maintainer: the interface for a maintainer shall ease the isolation of any malfunctioning system in MFAA. For the maintainer a set of test scripts is available to prove the correctness of MFAA. Fault reports should be simple and indicate the malfunctioning at unit/board level. In general, a pass/fail result should be sufficient for this purpose.
- Expert user: the interface for an expert user gives access to the lower detailed levels of the system. An expert user can check the system at a more detailed level and can set or inject known stimuli in the digital signal chain and monitor the output at several stages for fault analysis. This mode is also used for commissioning and verification of the MFAA system and subsystems. Also there is a set of test scripts available for the expert user to prove the correctness of MFAA at several stages.
- Triggerer: via this interface external triggers are received from other radio telescopes or transmitted to other telescopes. If applicable these triggers can be used to change the observation instantly to observe another source or for freezing data for example. This is indicated as a separate interface because the response time has a real-time nature.

- Scheduler: via this interface all the resource claims of the MFAA system over time are communicated.

### 6.1.5 INFRA-SA to MFAA Interface

The INFRA-SA to MFAA interface can be grouped in several underlying interfaces:

- Terrain: this consists of the actual field wherein the antennas need to be installed. Typical interface requirements such as field flatness, size of the terrain, soil, presence of obstacles e.g. mountains amongst others apply to this interface. Also the accessibility of the MFAA equipment by roads or other means belong in this category.
- Building: the building interface is primary meant for the required signal processing systems necessary for MFAA to fulfil part of its function. These could be small shielded cabinets (huts) close to the antennas or a central building handling for example the whole core and stations out to many kilometers.
- Cooling: also the cooling interface primary applies to the required signal processing in the building or smaller huts. The nature of the interface depends on a number of design choices which need to be made later in the development process.
- Power: the power distribution to MFAA and within MFAA needs to be agreed and decided on by these interface requirements. The main power from the grid or other sources need to be distributed to MFAA in the field, to the huts and/or to a central building close to the core.
- Time/Frequency reference: a time/frequency reference is required to synchronize all signal paths with each other and to provide a time stamp for the data such that data distribution from there on can be transported asynchronously. For AAMID it is proposed to provide these communication links via INFRA-SA because there might be optimizations possible to combine this with the power or network distribution. It falls under the infrastructure category and if trenches need to be dug, it is cost effective to combine these infrastructural functions.
- Network: this interface is necessary to transport the data to the central signal processor but also for distributing the control signals to the MFAA system.

### 6.1.6 MFAA to Central Signal Processor Interface

Although the physical connection is realised by the infrastructure interface, the higher levels of the OSI layers will be defined by this interface using standard interfaces. Typically, this consists of requirements concerning datarates, packet definitions, protocols, etc. Also, the amount of tolerable error loss will be defined within this interface. This has also a close connection to the infrastructure interface.

## 6.2 Functional System Decomposition

Some or all of the following functions shall be performed by the MFAA system:

- Receive electromagnetic signals
- Amplify signals
- Condition signals for beamforming
- Beamform, in the analogue domain, multiple signals
- Condition signals for transport
- Transport analogue signals
- Receive transported signals
- Condition signals for digitization

- Digitize signals
- Channelize signals
- Calibrate the signals
- Beamform multiple digital signals
- Calibrate beams
- Transmit beam data to central signal processor
- Report estimated beam shape and performance information
- Control and Monitor the MFAA system
- Distribute power and clock to relevant items
- House MFAA related functions

## 6.3 Functional Architecture

The functional architecture of one logical MFAA station is depicted in Figure 8. Not all functions are depicted specifically in the diagram for clarity. Since the baseline length need to vary from experiment to experiment (Section 5.6) it may be advantageous to adopt a centralised concept (see Section 9.5) over a distributed concept. Figure 8 presents a generic architecture without implying the physical implementation and location of the functions. Also the use of an analogue beamformer and location of the ADCs are design decisions still to be taken in a later phase.

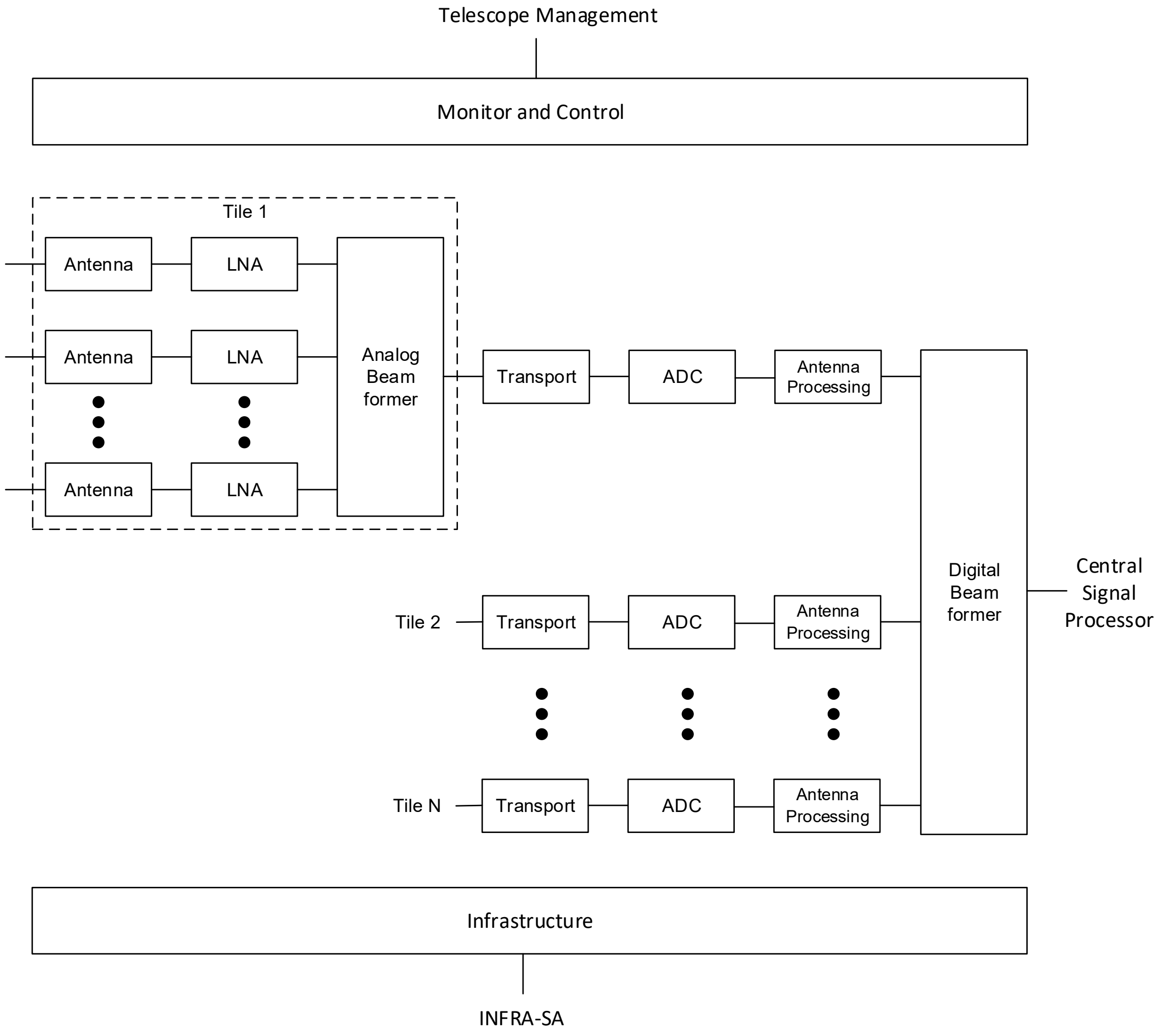

**Figure 8: Generalized functional architecture of a logical MFAA station composed of N tiles, note that the analogue beamformer is a design decision and that all antennas may be directly digitised and digitally beamformed.**

The antennas at the left of the diagram receive the electromagnetic signals. Accordingly, a Low Noise Amplifier (LNA) amplifies the signal. Both the antenna and LNA need to be designed in close interaction in order to find a design optimum in terms of (1) noise contribution, (2) power and (3) cost. The latter also includes the cost of manufacturing the antennas and to assemble them in the field. The signal is conditioned with additional amplifier stages and filtering before the optional analog beamformer. The main operation of the analog beamformer is to combine a subset of the antennas, further referred to as tile in order to "optically" select a portion of the sky. Therefore, this function reduces the antenna Field of View (FoV) resulting in data reduction. An all-digital architecture will be discussed in Section 10.2. The analog signals are transported to a processing facility where the digitization and digital processing takes place.

The transported signals are received and conditioned before A/D Conversion (ADC). Conditioning consists typically of filtering the Nyquist band and amplifying the signal such that the full ADC range is used. In the ADC the analog signal is sampled in time and quantized in amplitude resulting in digital signals.

The main role of the signal processing system is to calibrate the incoming signals across the bandpass and to combine signals from multiple antennas or tiles to form the station beams, wherein a station is defined as a group of antennas or tiles. The antenna processing presented can consists of time delays or a filter bank dependent on the signal processing architecture chosen. In the beamformer the digital signals of all antennas are combined to form a beam on the sky. By that another reduction of the FoV is made. However, by generating multiple digital beams the optical FoV of the tiles/antennas can be used effectively. The FoV of all station beams together is a trade-off parameter with cost of the digital processing facility and post processing capability. The station beam data is finally transported to the central signal processor.

The order of the functions and blocks in Figure 8 may vary, dependent on the adopted architecture, a number of options are discussed in Section 9. Since, many of those trade-offs are dependent on technology advances such design decisions should be postponed as long as possible. On the other hand, the continuous development and evaluation of systems and prototypes will ensure that the right choices are made and that a rapid increase of maturity (TRL) becomes possible, also for new technologies.

Monitoring and control is necessary to set the weights of the beamformers, attenuator settings and health monitoring of the antenna systems in the field. Furthermore, status information from the underlying hardware and signal chain is interpreted and sent up to higher control layers, if required. The monitor and control is also responsible for the calibration of the individual signal paths, based on data received by MFAA and sent to the monitor and control system.

Infrastructure within MFAA is responsible for the power, clock and data distribution within MFAA and therefore also in between stations.

# 7 Design Space Exploration

In this section a number of trade-offs are presented to explore the design space. The main goal of this exploration is to meet the requirements as presented in [RD-1]. In this section no particular mapping of functions to a physical architecture is assumed.

## 7.1 Array configuration

Antennas can be configured in several ways. Table 2 lists a number of properties for four array configurations, which is a combination of their density (dense/sparse) and their orientation (regular/irregular). Previous examples of radio astronomy aperture array configurations are:

- Dense regular: EMBRACE
- Dense irregular: no examples
- Sparse regular: LOFAR HBA, MWA
- Sparse irregular: LOFAR LBA, SKA1-Low (LFAA)

For MFAA two configurations are investigated: dense/regular and sparse/irregular. The dense irregular array configuration is practically impossible due to the size of the antenna elements, while the sparse regular arrays are not considered because of the presence of large grating lobes.

Two configurations are feasible for MFAA. A configuration wherein the antennas are more separated: sparse/irregular, and a configuration wherein the antennas are packed closely together dense/regular. Formally the term sparse and dense is not correct, since it should relate to a certain frequency. An array can operate in the dense regime or sparse regime. Assuming a pitch in between the antennas of $\lambda_p$ corresponding with a frequency $f_p$ then a design is in the

- Sparse regime for $f > f_p$
- Dense regime for $f < f_p$

In this document the term sparse array is used when the sparse regime is larger than the dense regime for the frequency band of interest (450 – 1450 MHz).

| | | **Regular** | **Irregular** |
|---|---|---|---|
| | Sidelobes | Lowered by gain taper | Lowered by space taper |
| **Dense** | Grating lobes | No | |
| | Receiver temp | Smooth (angle, freq) | |
| | Effective area | Constant over frequency, smooth over angle | |
| | Element patterns | Depend on position | |
| | Station diameter | Smaller for a given sensitivity | |
| **Sparse** | Grating lobes | Few high ones | Many low ones |
| | Receiver temp | Smooth (angle, freq) | Less smooth (angle, freq) |
| | Effective area | Varies with $\lambda^2$ | Varies with $\lambda^2$ |
| | | Not smooth (angle, freq) | Smooth (angle, freq) |
| | Element patterns | Constant for most elements | Depend on position |
| | Station diameter | Larger for a given sensitivity | |

**Table 2 Array types and their properties**

### 7.1.1 Dense regular arrays

In a dense array the antennas are closely packed. The pitch is chosen such that a first grating lobe enters the band at the top end of the band for the maximal scan angle. Due to the regularity of the array the mutual coupling in between the antennas is the same for all antennas, except for the boundary antennas. This makes the station beam pattern very stable, smooth and predictable. The station beam pattern is basically defined by the station geometry instead of the individual antennas.
Advantages:

- Constant A/T for higher frequencies

- Station beam pattern smooth and predictable with frequency and scan angle
- Low far outside-lobes

Disadvanges:

- Lower A/T for lower frequencies, due to increasing sky noise
- A curved A/T performance, which only gives nominal performance in the centre of the band
- A large number of antennas is required, since the pitch is driven by the highest frequency and the antennas are smaller

A sensitivity curve as a function of frequency for a dense array is depicted in Figure 9, wherein the sensitivity at zenith is set to 10,000 $m^2/K$ at 1 GHz. The gray scale indicates the sensitivity decrease with the scan angle as parameter. An average sensitivity for scan angles between 0 and 45 and 0 and 60 degrees is shown as well.

Since the antennas in a dense array are closely packed the station beam size is larger than for a sparse array because the station diameter for a given sensitivity is smaller. In [RD-12] more detailed information is available.

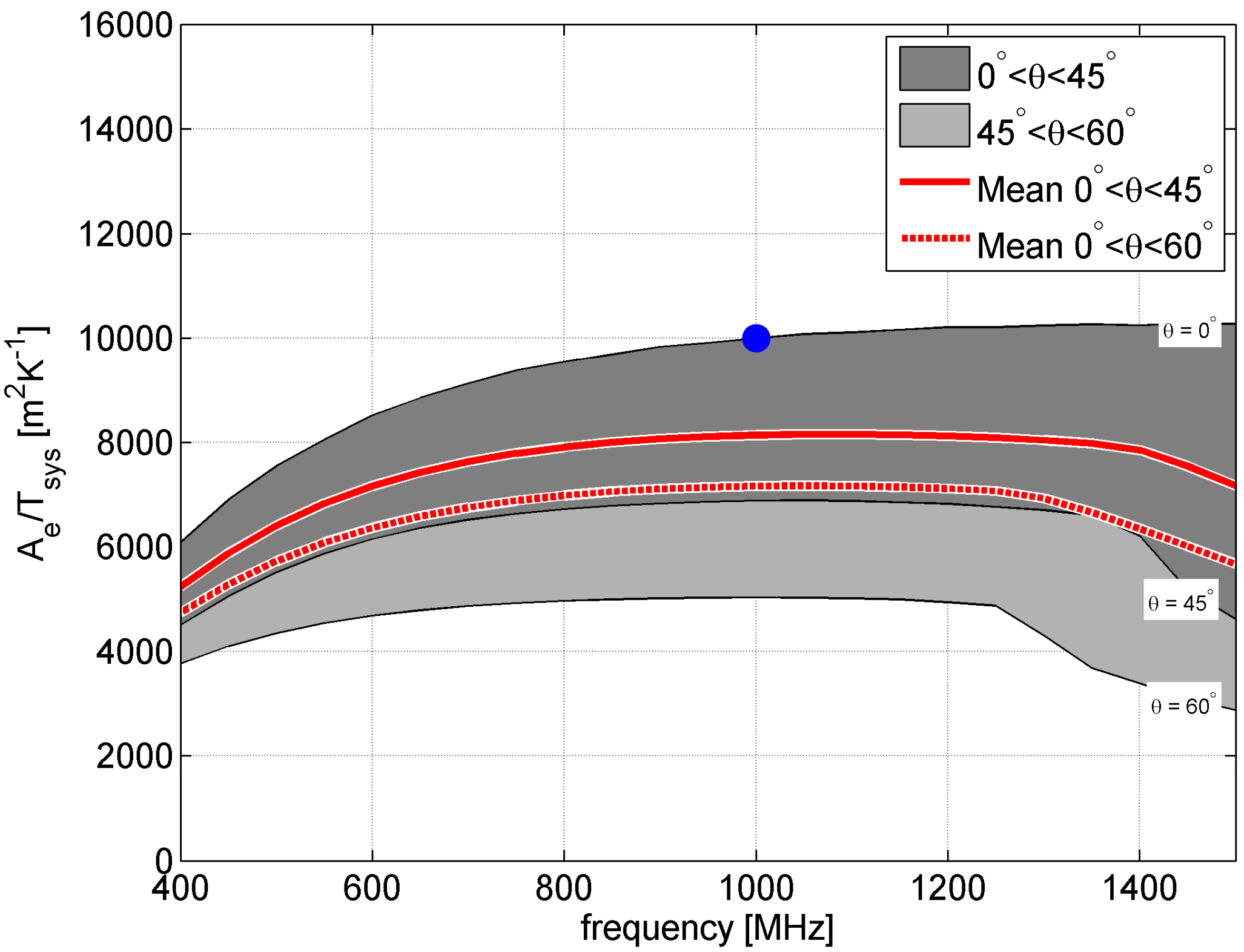

**Figure 9. Modelled sensitivity of a regular dense array, normalised to 10,000 $m^2K^{-1}$ at 1 GHz.**

### 7.1.2 Sparse irregular arrays

In a sparse array the antennas are spaced further apart than λ/2. In the design presented in Section 10.2 an antenna pitch of 35 cm was assumed. By using a sparse array, signals with a frequency larger than the pitch frequency are under-sampled in the spatial domain causing grating lobes to appear. A way to reduce or scramble the grating lobes is by using an irregular array.

Advantages:

- High A/T for lower frequencies
- Fewer antennas required for the same performance at lower part of the band

- Can have antenna gain in sparse regime for increased sensitivity at zenith

Disadvantages:

- A/T drops with $f^{2}$ profile within the band
- Station beam model less smooth and predictable as within a dense array
- Smeared grating lobes within the band

A sensitivity curve as a function of frequency for a sparse array is depicted in Figure 10, wherein the sensitivity at zenith is set to 10,000 m$^2$/K at 1 GHz. From this curve the superior sensitivity for the lower frequencies can be seen at the expense of sensitivity for higher frequencies. The curves also show a more rapid decrease in sensitivity as function of scan angle compared with Figure 9.

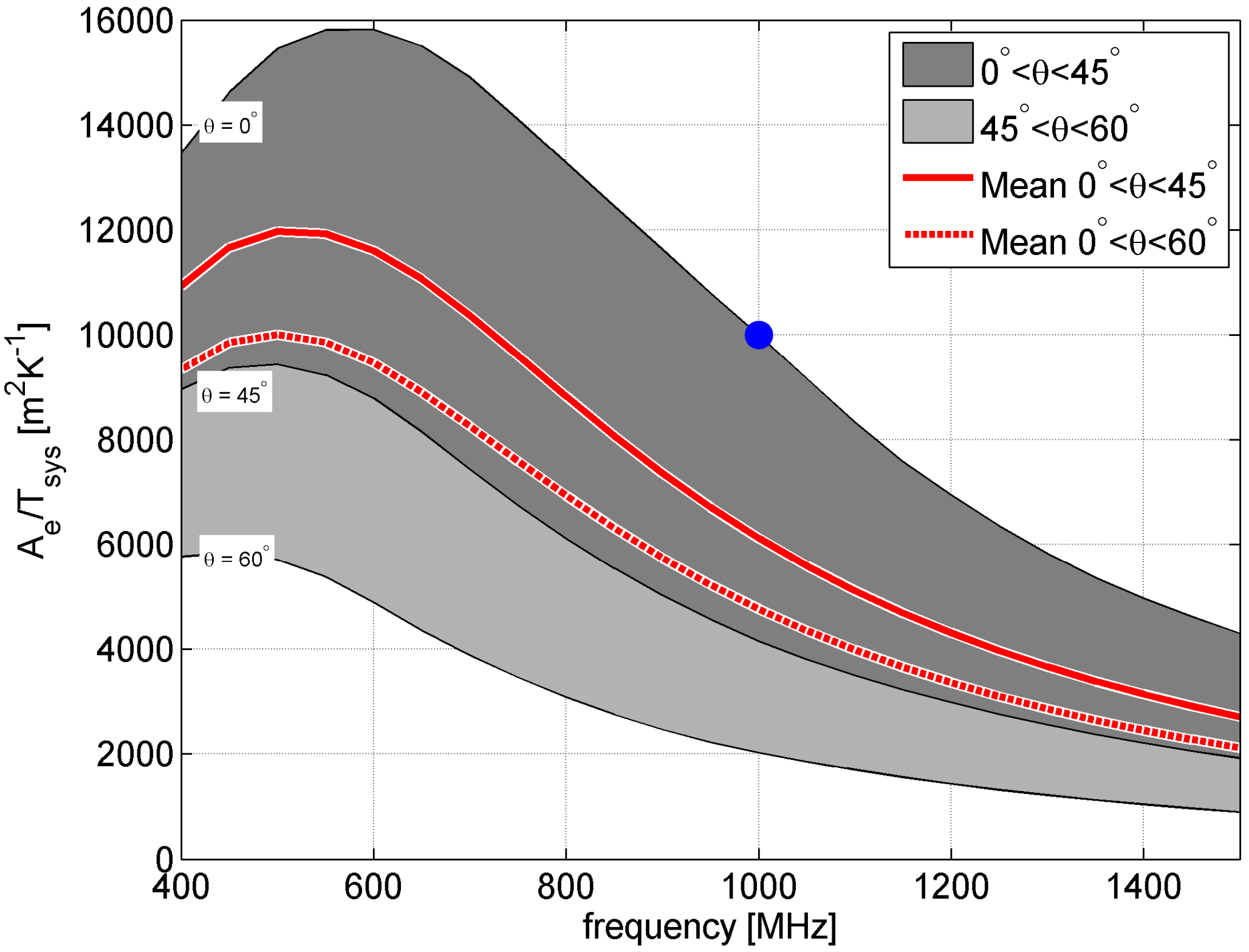

**Figure 10. Modelled sensitivity of a sparse irregular array, normalised to 10,000 m$^2$K$^{-1}$ at 1 GHz.**

An advantage of a sparse array is that more forward gain in the zenith direction can be realized, at the expense of gain in other directions. The antenna gain integrated over the full sky is a constant. Therefore, if an antenna or antenna array has more forward gain at zenith the gain from zenith to the horizon drops off with a higher slope. Typically, the antenna gain drops with cos($\theta$), with $\theta$ the scan angle. For an antenna with more forward gain, the antenna gain drops with cos$^2$($\theta$) or possibly even more. Figure 11 depicts the sensitivity as function of frequency for a sparse array, wherein the black curve assumes a cosine element pattern and the red curve a squared cosine element pattern. The dashed curves display the sensitivity as function of scan angle. In this example it can be seen that the sensitivity of individual low gain and high gain antennas are similar at 45° scan angle with the benefit that the high gain antenna has significantly more sensitivity at zenith. The reduction at 60° scan is minimal. Low gain towards the horizon is a benefit in reducing terrestrial RF interference.

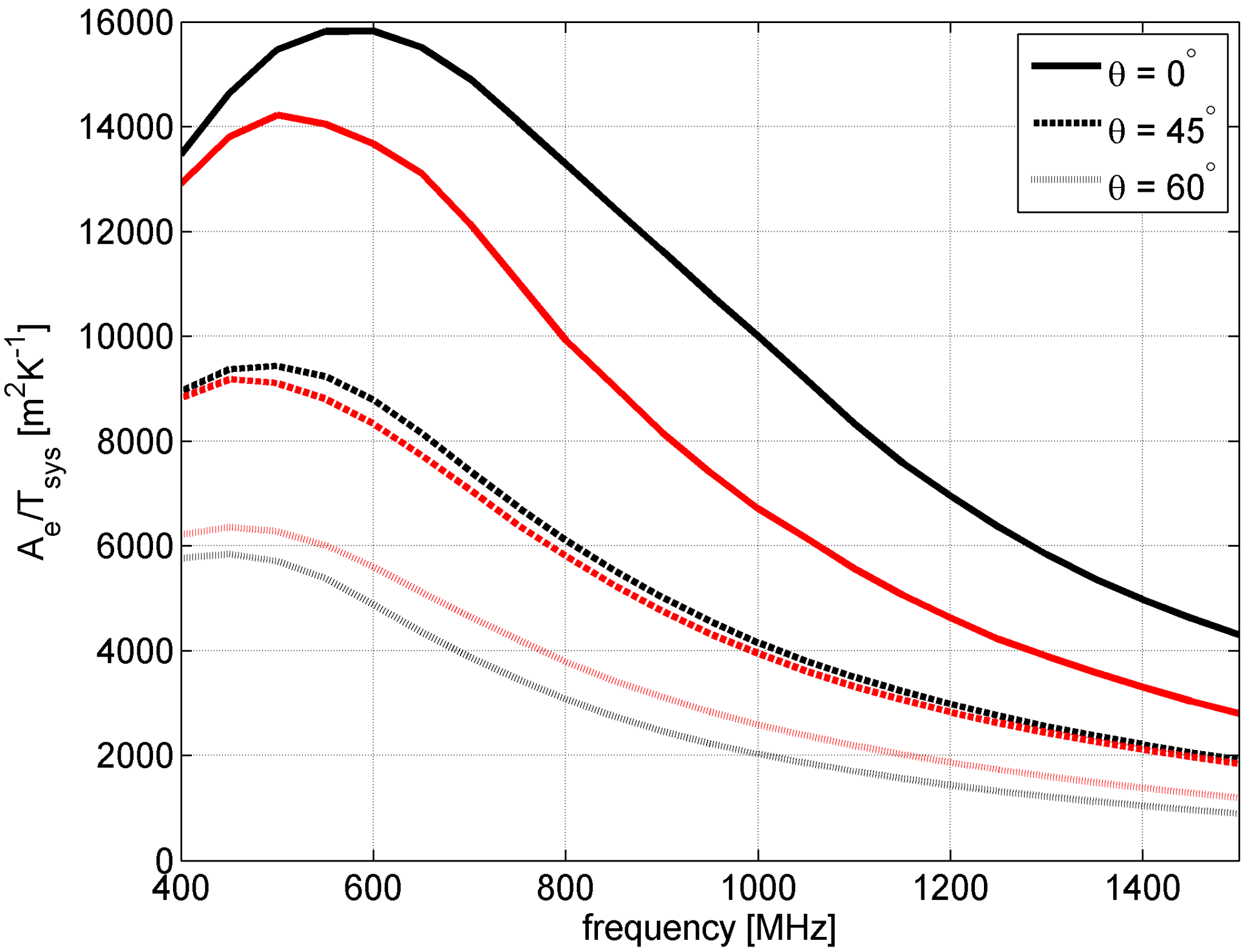

**Figure 11. Sensitivity as function of frequency of a sparse irregular array at three scan angles for a cosine element pattern (red curves) and cosine² pattern (black curves)**

## 7.2 Technical Assumptions

In this section the main technical assumptions are discussed which are relevant for the rest of the design space exploration.

### 7.2.1 Receiver Temperature

A target receiver temperature of 30 Kelvin is assumed for the full bandwidth of 450-1450 MHz. The system temperature is calculated as $T_{sys}=T_{sky}+T_{rec}$ wherein the sky temperature is defined is in line with [RD-6] as

$$T_{sky} = \left(\frac{c}{0.2008f}\right)^{2.55} + \left(\frac{f}{f_0}\right)^{1.8} + T_{bg}$$

with *c* the velocity speed of light, *f* the frequency, $f_0$=1 GHz and $T_{bg}$=2.7 K the back ground noise. The system temperature assuming a Trec of 30 K [RD-12] is shown in Figure 12.

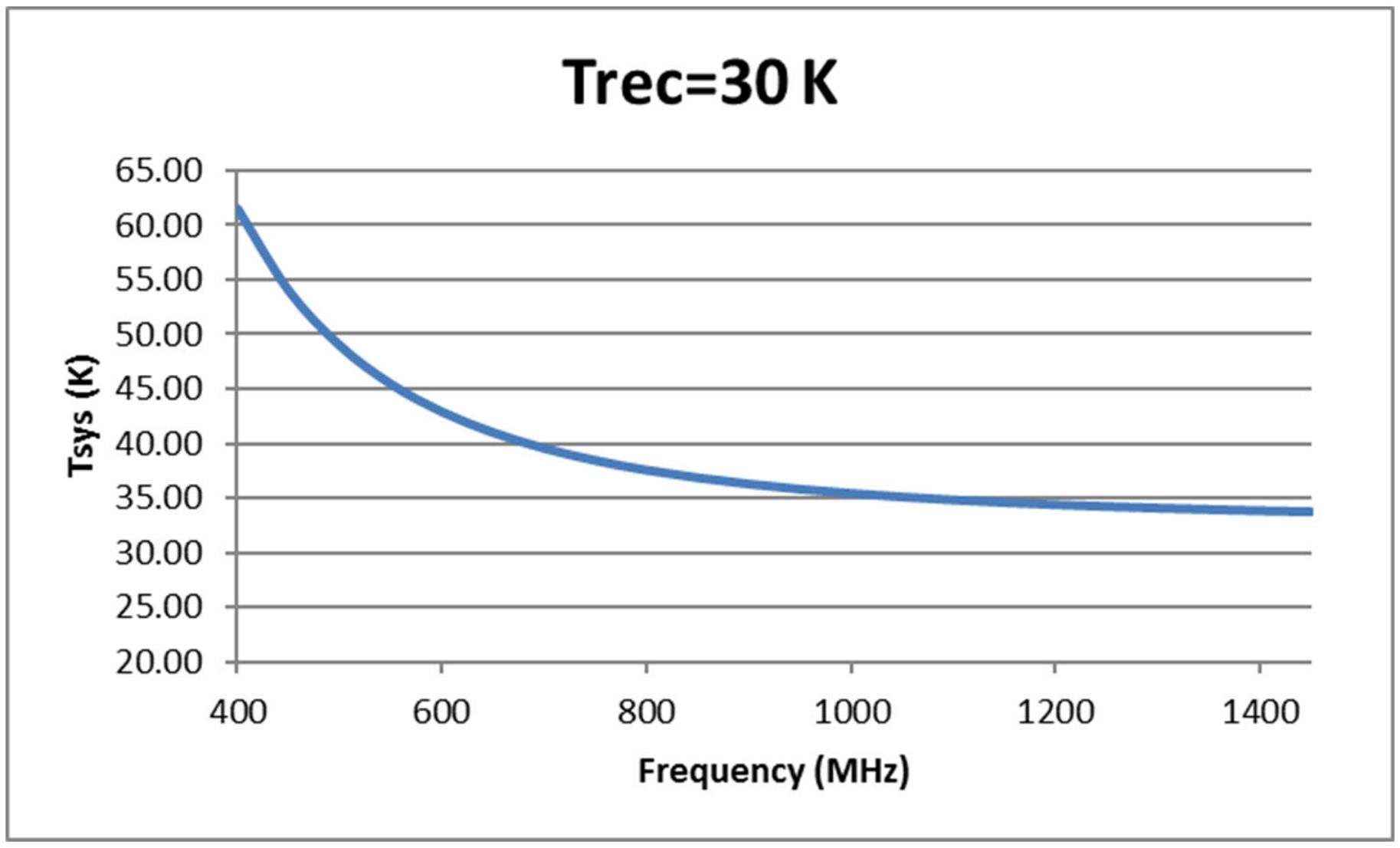

**Figure 12: System noise temperature as a function of frequency**

### 7.2.2 Effective Area

The sensitivity of MFAA is linearly proportional with the total collecting area available. A measure of the collecting area is defined as effective area. For the dense/regular antenna designs presented, an antenna spacing is assumed at a pitch frequency $f_p$ of 1.2 GHz. This frequency is chosen such that a grating lobe starts to become visible only at the top of the band at 1.4 GHz, given a maximal scan angle of 45 degrees relative to zenith. This results in an antenna pitch *d* of 0.125 m.

The effective area of one antenna element $A_{el,eff}$ at zenith is approximated as:

$$A_{el,eff} = d^2$$

In Figure 9 the effective area profile is shown. A lower $f_p$ results in fewer antennas and is a direct cost reduction. However, this comes at the expense of grating lobes.

## 7.3 Technical Designs for Regular Dense Arrays

In this section a number of design trade-offs are presented. In the past there have been a numerous amount of iterations from the science requirements to a design, giving a certain cost cap, and back to the science again. The goal of this section is to show a number of parameters which can be traded-off, resulting in the same overall cost cap. It is meant to show the choices which can be made and their impact on the performance. The absolute performance measures which can be achieved given a certain budget is presented in [RD-13] and for two system concepts in Section 10.3.

### 7.3.1 Tile Size Trade-off

Reducing data at the front of the system can be advantageous from a cost perspective. For example if the number of antennas combined in Figure 8 is maximized for the first stage beamformer the system cost for the rest of the system may be reduced (need fewer signal paths to transport and digitize). Consequently, since we assume the same costs for the designs, more area can be created to increase the sensitivity.

According to [RD-8] the design can be optimized for cost when the tile size or base receptor element as is used in [RD-8] is tailored to the required optical FoV. This minimizes the number of signal paths. FoV, also indicated with ΔΩ, in steradians is defined in this document as:

$$\Delta\Omega = \frac{\pi}{4} \cdot \frac{\lambda^2}{D^2}$$

wherein $\lambda$ is the wavelength and $D$ the diameter of a tile. This can be converted to square degrees by multiplying by $(180/\pi)^2$. The upper limit of the tile size (assuming the beamformer can handle the dimensions of the tile) is primary driven by the optical FoV requirement. In Figure 13, three designs are presented with different tile sizes. The tile sizes used are 2 $m^2$, 1.125 $m^2$ and 0.5 $m^2$ giving an optical FoV at 1 GHz of respectively 58 $deg^2$, 103 $deg^2$ and 232 $deg^2$. From this can be seen that for similar cost either a large tile with a small optical FoV but a large A/T or a smaller tile with a larger optical FoV and smaller A/T can be built.

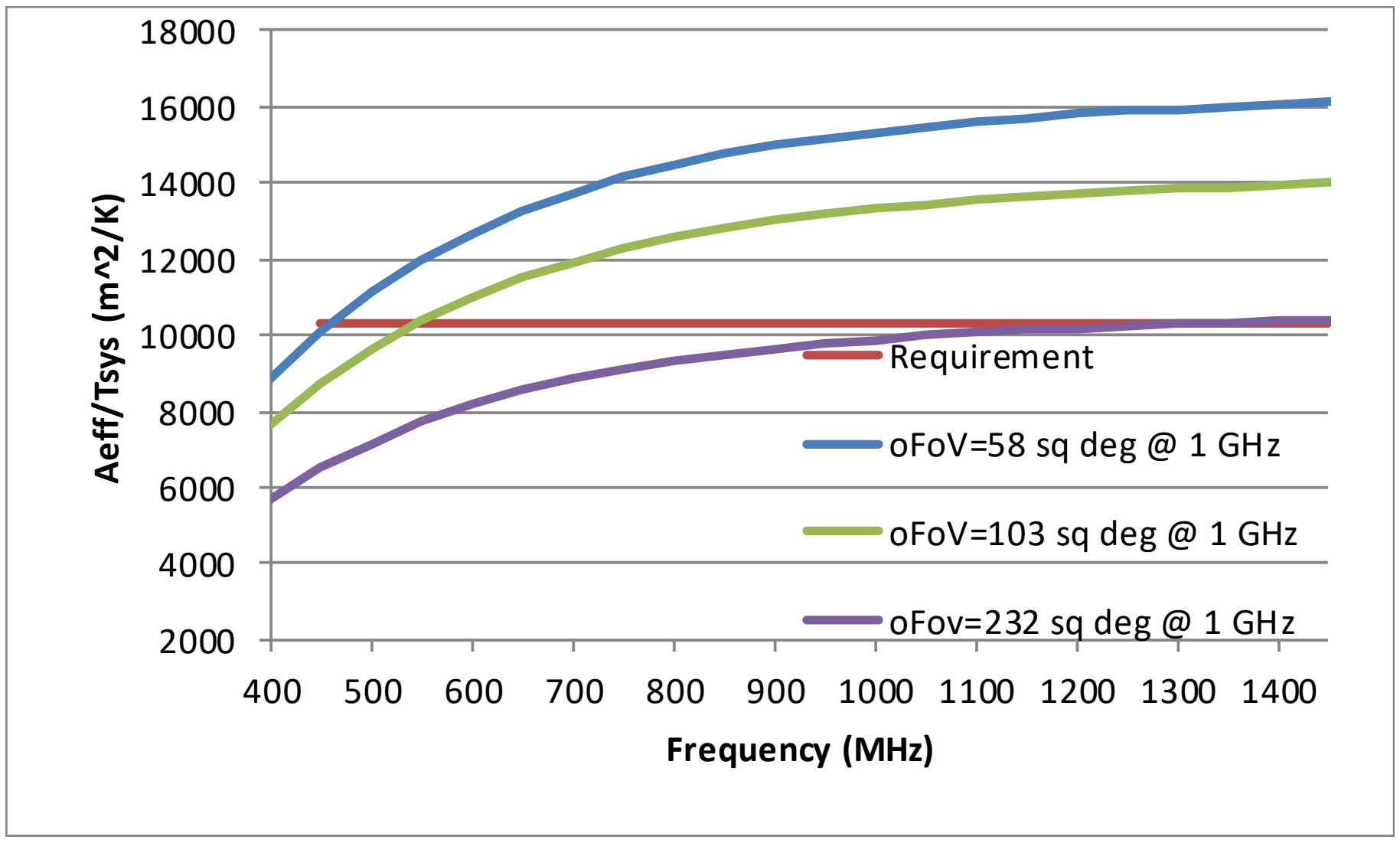

**Figure 13: A/T performance as a function of frequency at zenith**

Figure 14 shows the optical FoV profile as function of frequency which can be achieved for the various designs. Those FoVs can be achieved only if sufficient bandwidth is traded-off.

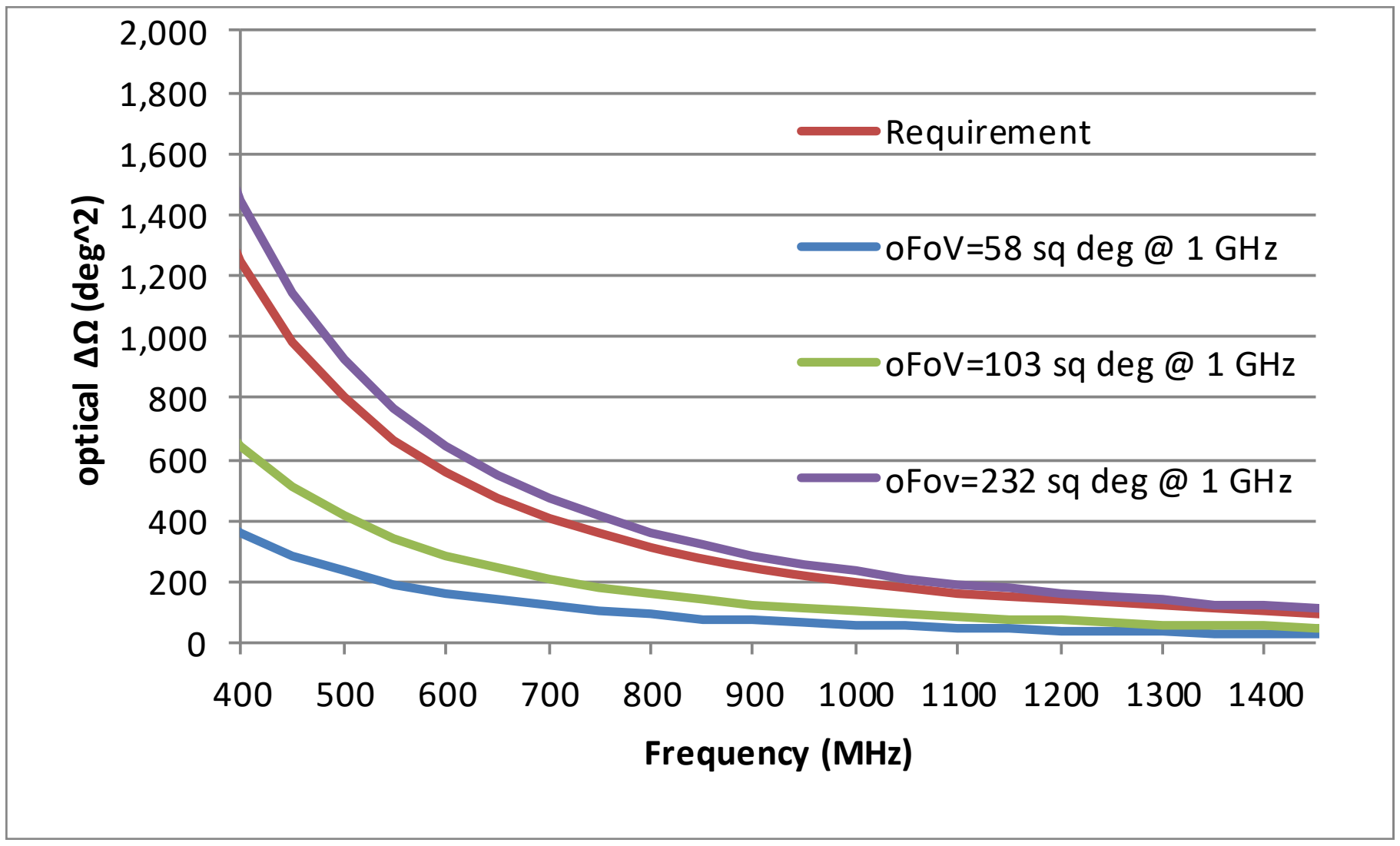

**Figure 14: Optical FoV for several tile sizes**

An alternative way to increase the optical FoV is to generate an extra analog beam from the beamformer. The disadvantage is that the number of signal paths and therefore signal processing is doubled. However, the optical FoV is increased by that technique. Moreover, there is considerable extra flexibility since both beams can be steered independently from each other. The generation of at least two optical FOVs is a requirement for scientific flexibility.

To achieve a required optical FoV of 200 $deg^2$ at 1 GHz, a tile of 1.125 $m^2$ could be used delivering 103 $deg^2$ for each analog beam. By using two analog beams the requirement will be met. Another solution is to use a tile of 0.5 $m^2$, delivering an oFoV of 232 $deg^2$.

The maximum survey speed curves for the designs are show in Figure 15. Survey speed is defined as

$$SS = \left(\frac{A_{eff}}{T_{sys}}\right)^2 . \ FoV$$

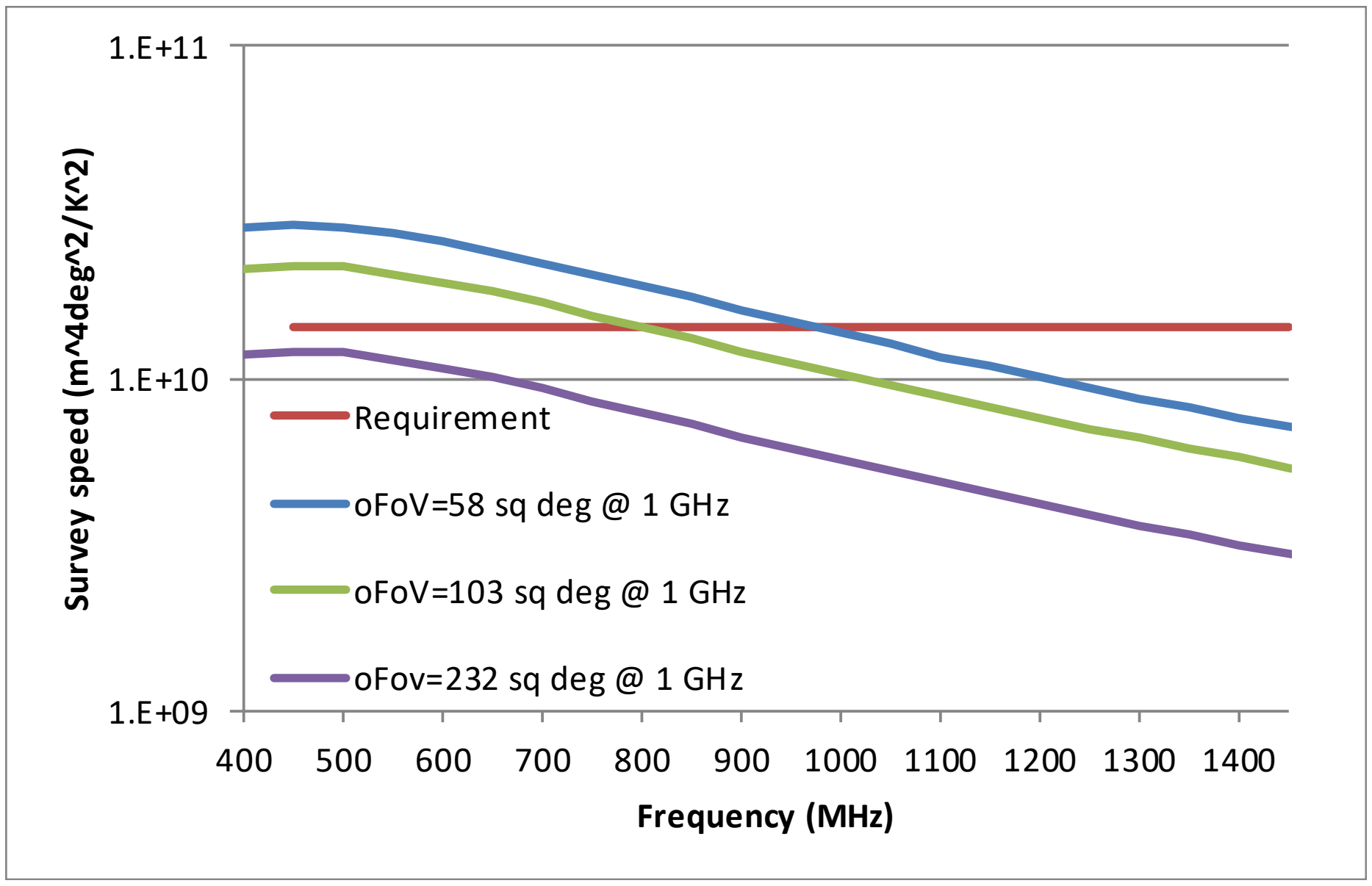

**Figure 15: Survey speed for a tile of 0.5, 1.125 and 2 m² at zenith**

A summary of the design parameters for the presented designs is listed in Table 3.

| | oFoV=58 sq deg, BW=500 MHz | oFoV=103 sq deg, BW=500 MHz | oFoV=232 sq deg, BW=500 MHz | Unit |
|---|---|---|---|---|
| Number of stations | 250 | 250 | 250 | |
| Number of pol.'s | 2 | 2 | 2 | |
| Number of elements/tile | 128 | 72 | 32 | single pol. |
| Number of tiles/station | 1089 | 1681 | 2809 | |
| Number of elements all stations | 35 | 30 | 22 | M |
| Number of beams/station | 1089 | 945 | 702 | |
| Input bandwidth/signal path | 500 | 500 | 500 | MHz |
| Output bandwidth/beam | 500 | 500 | 500 | MHz |
| Antenna pitch | 0.125 | 0.125 | 0.125 | m |

**Table 3: Design parameters for the tile size trade-off**

### 7.3.2 A/T versus Processed FoV Trade-off

For this analysis a tile size of 1.125 $m^2$ is chosen, such that it complies with the optical FoV requirement. Two independent analogue beams are used, since at least two independent optical FoVs need to be created. Assuming the same cost, A/T can be traded with processed FoV. This results in the performance curves depicted in Figure 16 to Figure 18. The “A/T>” curve is a design which is optimized for A/T and therefore has a lot of collecting area. The “FoV>” curve represents a design with less collecting area, but with more signal processing to cover a larger FoV. The “Optimal” design presents a design which is in the middle of both. For the same cost a design can be chosen with either a higher A/T or higher processed FoV.

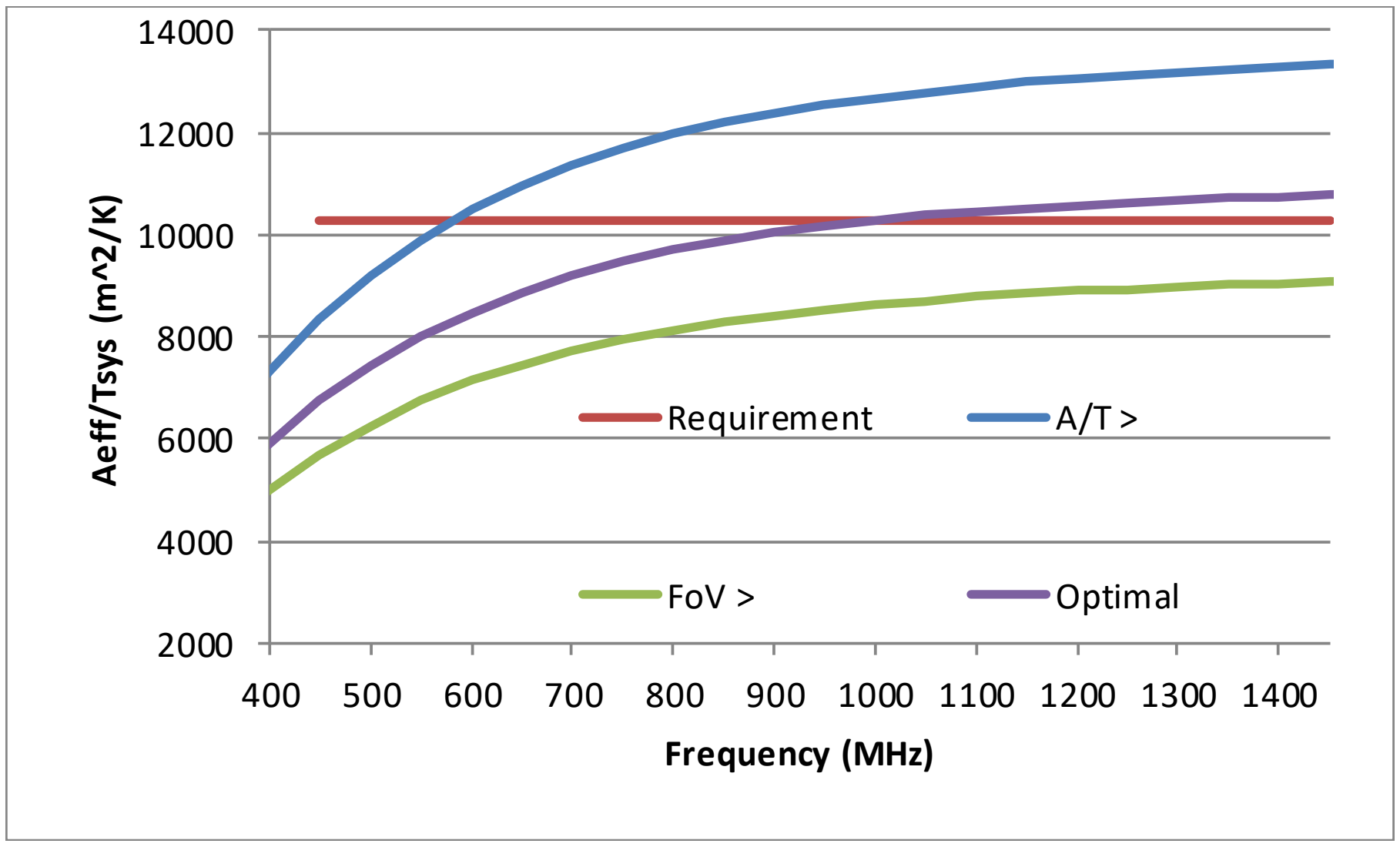

**Figure 16: Aeff/Tsys as a function of frequency at zenith for a design with large A/T, large FoV and an optimal design in between both**

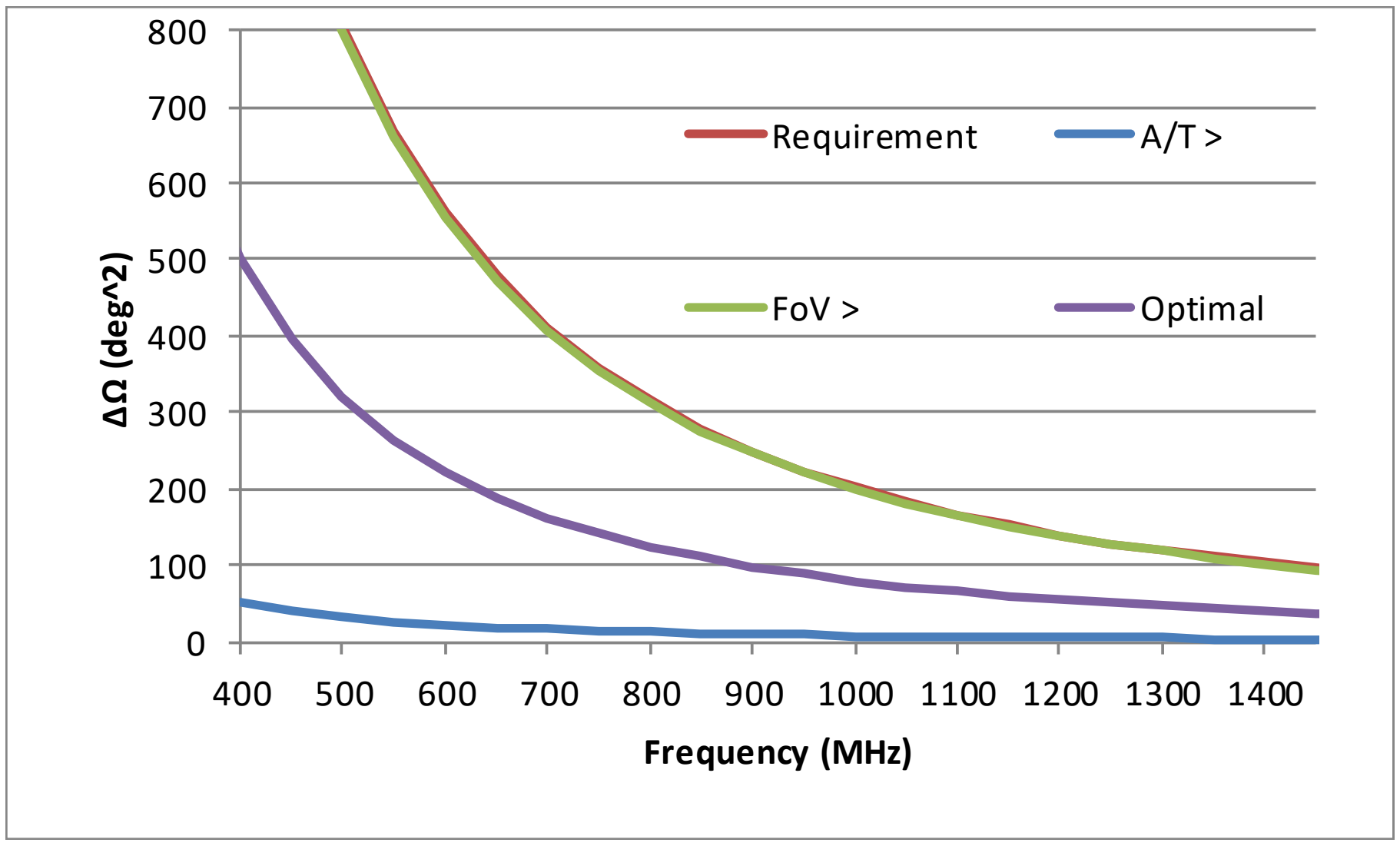

**Figure 17: FoV as a function of frequency at zenith for a design with large A/T, large FoV and an optimal design in between both**

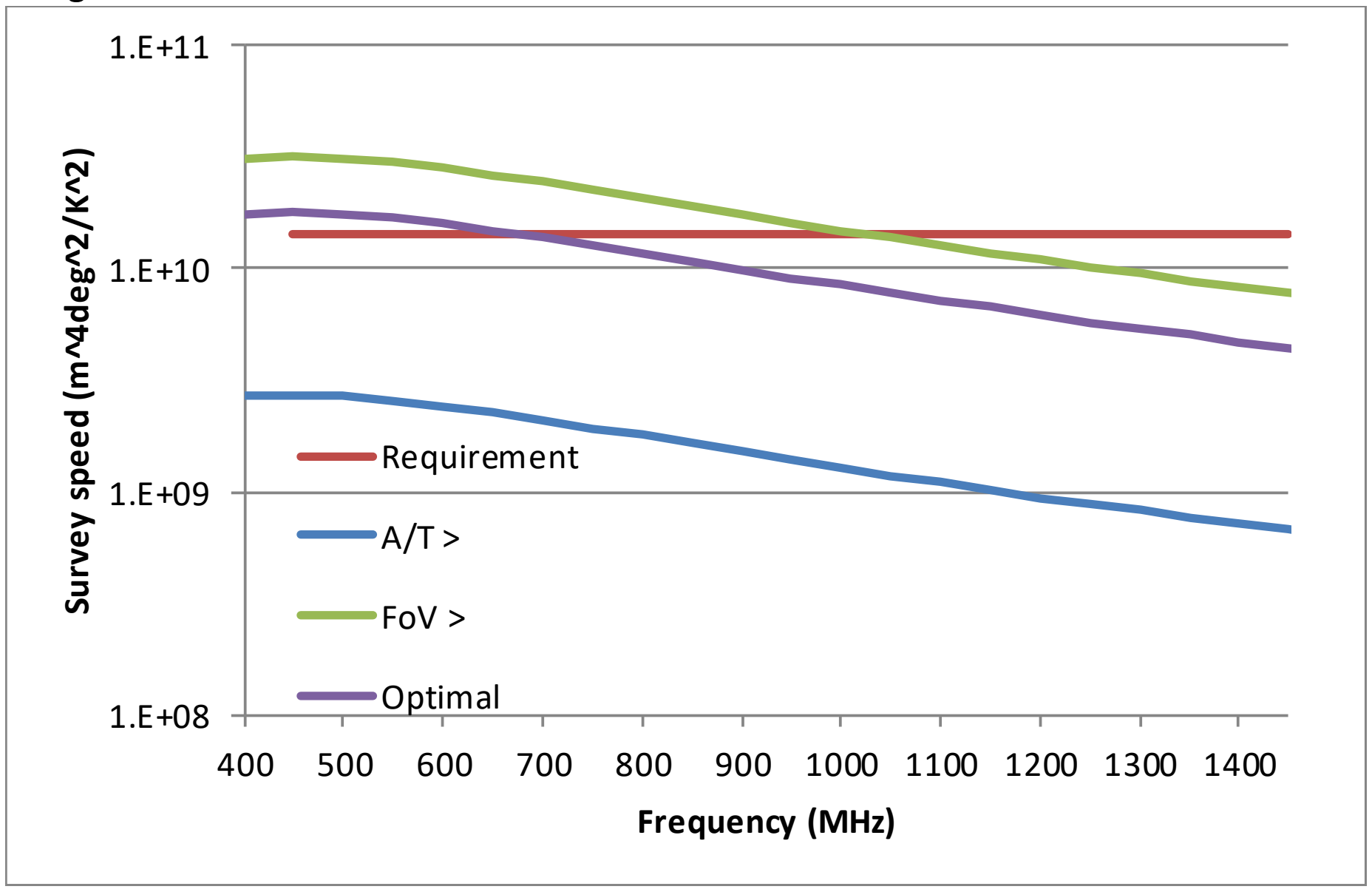

**Figure 18: Survey speed performance as a function of frequency at zenith for a design with large A/T, large FoV and an optimal design in between both**

The total assumed instantaneous bandwidth is 500 MHz. More processed FoV could be obtained by exchanging bandwidth for FoV as is discussed in Section 7.4.

A summary of the design parameters used for the presented designs is listed in Table 3.

| | A/T > | FoV > | Compromise | Unit |
|---|---|---|---|---|
| Number of stations | 250 | 250 | 250 | |
| Number of pol.'s | 2 | 2 | 2 | |
| Number of elements/tile | 72 | 72 | 72 | single pol. |
| Number of 1st stage beams | 2 | 2 | 2 | |
| Number of tiles/station | 1600 | 1089 | 1296 | |
| Number of elements all stations | 29 | 20 | 23 | M |
| Number of beams/station | 124 | 2112 | 1005 | |
| Input bandwidth/signal path | 500 | 500 | 500 | MHz |
| Output bandwidth/beam | 500 | 500 | 500 | MHz |
| Antenna pitch | 0.125 | 0.125 | 0.125 | m |

**Table 4: Design parameters for the A/T versus FoV trade-off (cost is the same for all designs)**

## 7.4 FoV versus Bandwidth Trade-off

In the designs presented so far the assumed instantaneous bandwidth was 500 MHz. Furthermore, the FoV profiles shown so far, all follows a natural profile which goes as $f^{-2}$. By channelizing the band in the digital domain flexibility to trade bandwidth for FoV. Suppose in total $N_c$ channels are created and $N_b$ beams are required for the full 500 MHz. A channel with a designated direction is here defined as beamlet. In total $N_c.N_b$ beamlets are available in the system. These could be arbitrarily spread over the bandwidth available. In the case we only use one tenth of the channels at a particular frequency, these can be used to increase the number of beams for the other channels by a factor of 10. In this way the FoV profile over frequency can be arbitrarily constructed, provided that the optical FoV is greater at all frequencies than the processed FoV. This is a major advantage of using aperture array systems. For example assuming a design that has a FoV of 80 deg$^2$ at 1 GHz ("Optimal design" from Figure 17) with a 500 MHz instantaneous bandwidth. By assuming the same instantaneous bandwidth any of the FoV profiles (or others) of Figure 19 can be constructed.

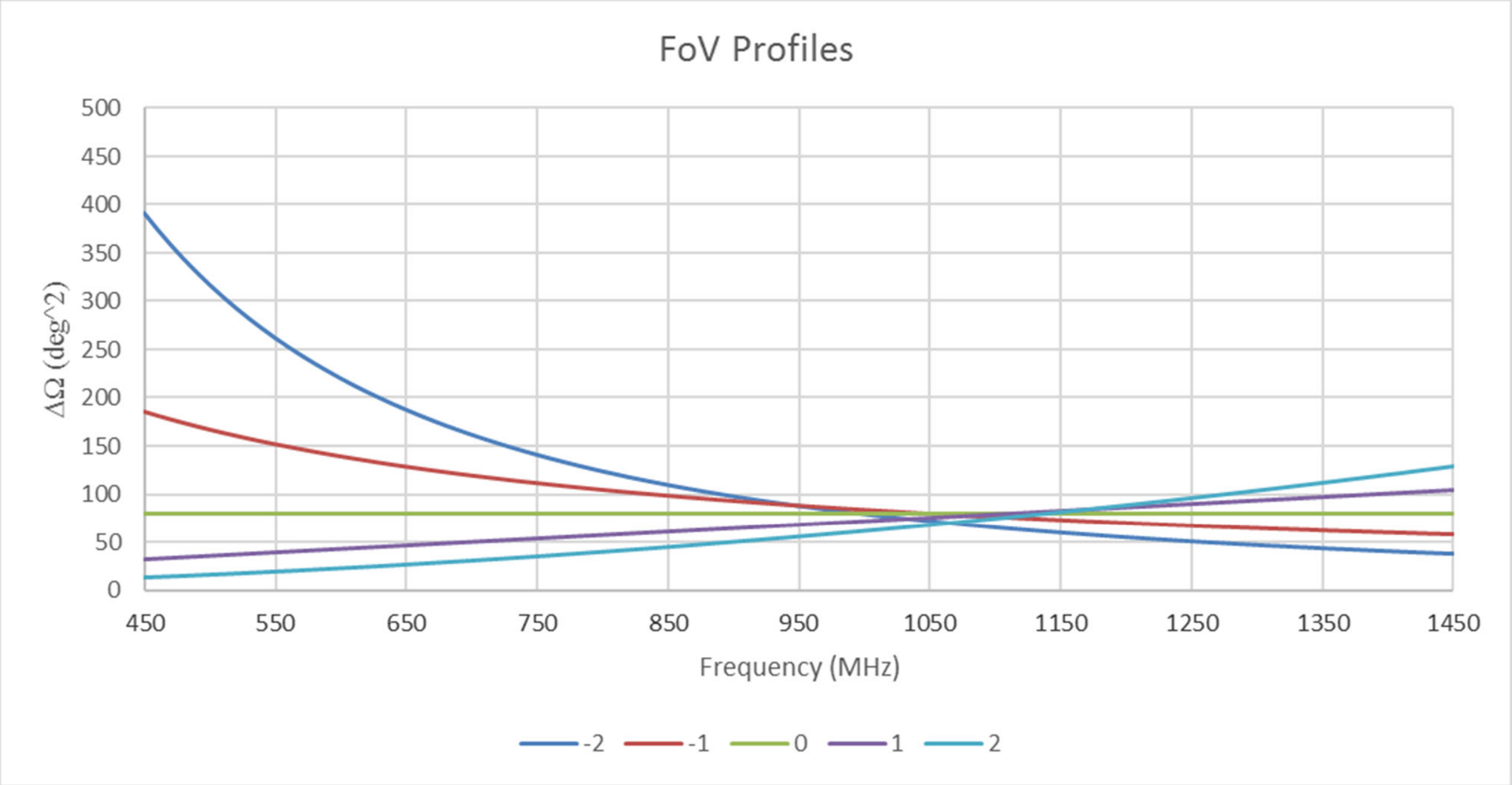

**Figure 19: FoV profiles $f^{-2}$, $f^{-1}$, $f^{0}$, $f^{1}$ and $f^{2}$ as a function of frequency for an instantaneous bandwidth of 500 MHz**

By trading bandwidth for FoV the total FoV can be enlarged up to the maximum optical FoV. For example, the FoV profiles assuming an instantaneous bandwidth of 200 MHz are depicted in

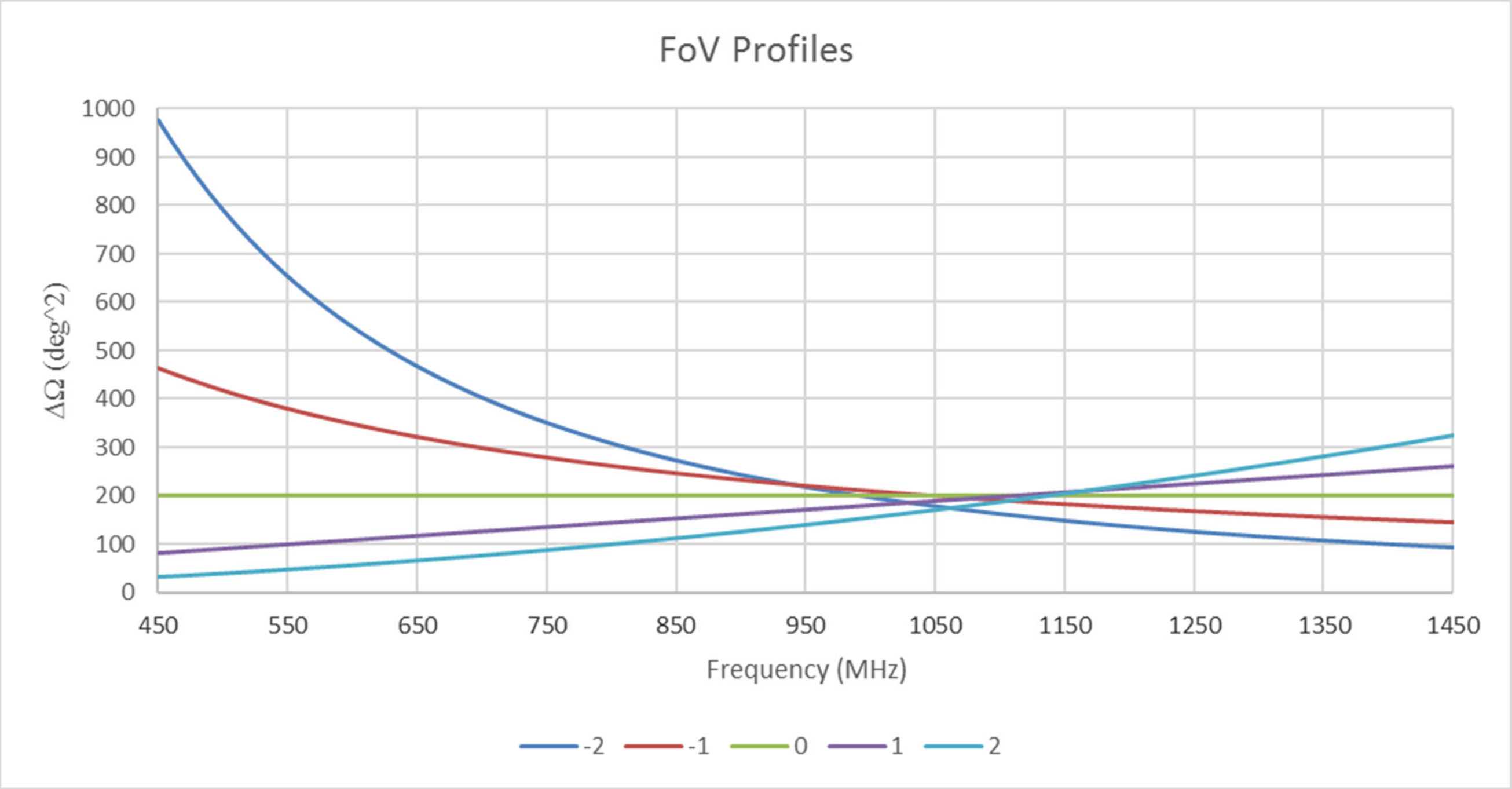

**Figure 20: FoV profiles $f^{-2}$, $f^{-1}$, $f^{0}$, $f^{1}$ and $f^{2}$ as a function of frequency for an instantaneous bandwidth of 200 MHz**

# 8 Calibration

The AAMID system will require calibration at multiple levels: within the station, the gain and phase difference between the individual receive paths need to be determined and corrected for and at array level, instrumental and environmental effects necessitate a calibration effort to compensate for those effects. Calibration at array level is not the responsibility of MFAA; this can only be performed at the consolidation of the signals by the post processing systems in the science data processor. This calibration will not be discussed in this document. Instrumental calibration, however, is the responsibility of MFAA and is essential to the basic MFAA requirement of providing high-quality, well-characterised beams to the correlator.

## 8.1 Calibration methodology

The approach to instrumental calibration is based on techniques which are developed for LOFAR and MWA and will continue to be refined in LFAA whereby some astronomical data is collected from all antennas and analysed to identify the gain and phase variations of the analogue signal chains. This is by taking the signals from groups of antennas and processing them to identify the amplitude and phase errors in the gain chain as a function of frequency. This enables the calculation of the coefficients to be used in the beamformer to compensate the gain and phase differences between receiver paths. At the time of a recalibration, the gains and phases should be set essentially accurately. The rate of calibration drift over time due to temperature and other effects then determines how long it takes before the errors in chains degrade the precision of the station beams to an unacceptable level.

The calibration process operates by frequency channel and is repeated for a sufficient number of frequency channels to interpolate across the entire passband. The data from each single frequency channel for a group of antennas are corrected for sufficient integration time to obtain a reasonable signal-to-noise ratio on bright astronomical sources to perform calibration of the channel. The errors found in the processing are used to calculate the calibration parameters of that channel for all the (groups of) antennas. The process is repeated at a cadence sufficiently short to ensure that the calibration of the tiles/antennas is accurate enough for the observations.

## 8.2 Calibration algorithms

The final algorithms that will be used by MFAA are yet to be determined and will depend on the details of the implementation. However, approaches are available and are being developed as noted above. Essentially there are the following techniques that are considered or developed, further options may be found:

- Pre-calibrate and use tables to set coefficients which depend on environmental characteristics, principally temperature. This can only work for relatively imprecise calibration over short distances e.g. short lengths of coaxial cable to reasonably stable amplifiers.
- Make a sky observation using signals from the antenna or beamformed tile. By using a station or reasonable group of antennas these can be correlated and the resulting signals used to calibrate on bright astronomical objects.
- Holographic techniques whereby a reasonably calibrated station can calibrate individual antennas in adjoining stations. This can be bootstrapped for improving calibration.

# 9 Physical Design Solutions

Currently there are number of sub-system concepts on the table, which deserves further study and trade-offs towards the PDR. Those sub-system concepts can be combined to generate a full MFAA system. In this section the sub-system concepts are further discussed.

At technology level also different types of technology are considered, such as the antennas. This is further discussed in the technology document [RD-12].

## 9.1 Analog Tile versus Digital Tile

The location of the A/D converter is a crucial choice and can result in an analog tile or digital tile. For the first concept a fully analog tile is assumed. The digitisation of the signals takes place in a central station cabinet. The analog signals can be transported via coaxial cable or RF over fiber. For the digital tile the A/D conversion is located near the antennas within the tile.

## 9.2 Analog Tile

The analog tile is shown in Figure 8. The antennas within a tile are combined with an analog beamformer and the electrical output is transported to a central processing location. There the electrical signal is converted into the digital domain.

Advantages:

- No clock distribution required in the field
- Digital signals close to the tile do not couple in the analog circuitry
- Low electrical power in the field

Disadvantages:

- Extra gain is necessary within the tile to compensate for the signal loss in transport
- Gain and phase differences are introduced by the long cable as a function of temperature

The analog tile concept has been demonstrated with EMBRACE [RD-5] for a frequency range of 500-1500 MHz.

## 9.3 Digital Tile

The second concept developed is shown in Figure 21. Within this concept the A/D conversion is located in tiles. This means that the A/D clock has to be transported to each tile.
Advantages:

- No extra analogue gain is necessary within the tile to compensate for the loss in the transport
- No errors are introduced by signal transport

Disadvantages:

- Clock distribution required in the field
- Digital signals close to the tile can couple in the analog circuitry
- More electrical power in the field

Designs for the subsystems of the digital tile are already in development as part of the Aperture Array Integrated Receiver (AAIR) work [RD-9].

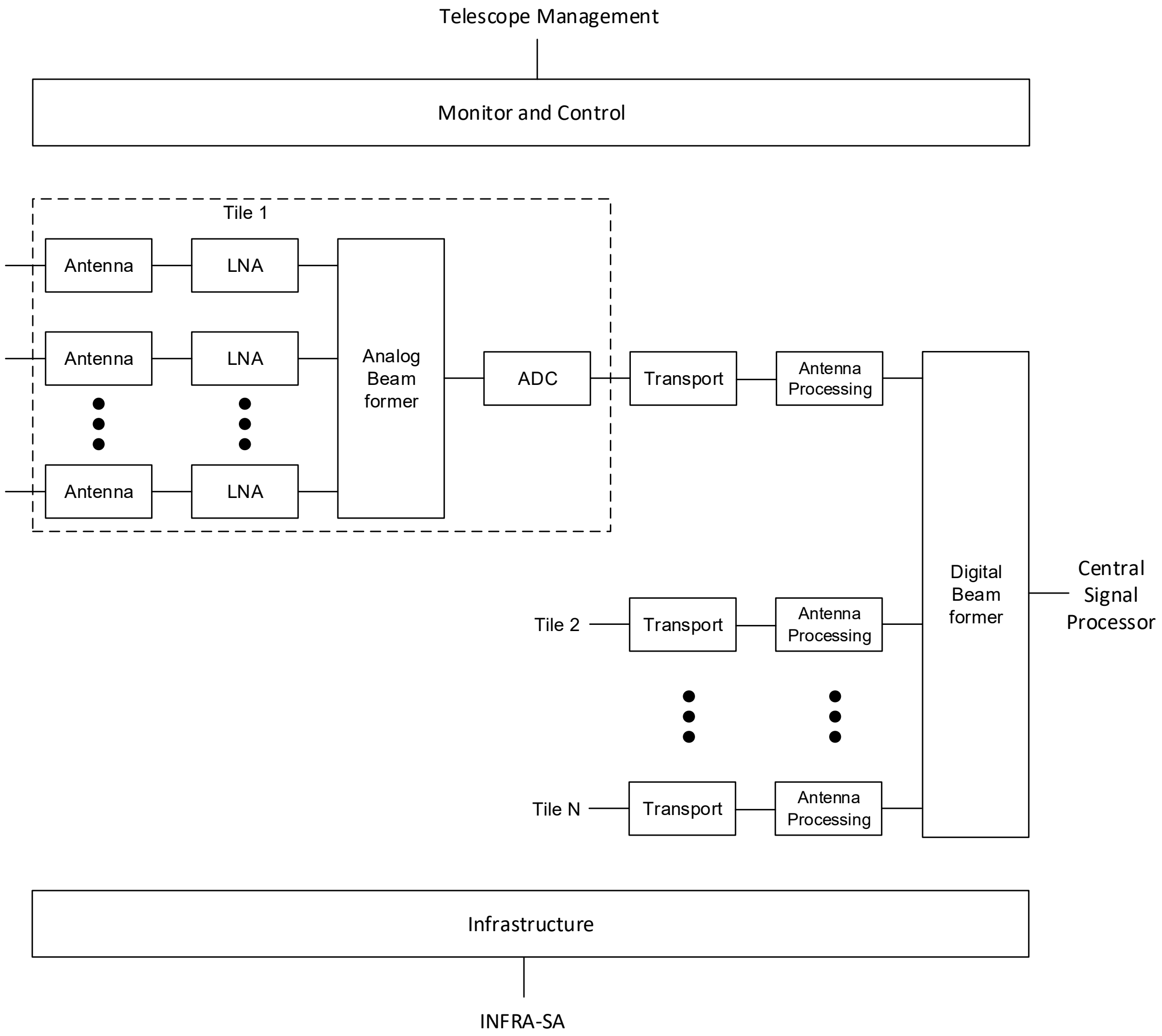

**Figure 21: Digital tile concept**

## 9.4 Analog Beamforming Compared to All Digital

The first stage beamforming can be done with an analog beamformer or by a digital beamformer. For the latter all antenna signals need to be digitized.

### 9.4.1 Analog Beamforming

The analog beamformer is the function to align the signals of individual antennas and thereafter sum these signals. This basically reduces the FoV. Although a time delay is required for this alignment, this can be approximated by a phase shift over bandwidths of more than 300 MHz.

In EMBRACE a dual stage beamforming technique is applied. The first stage beamforming of a group of elements is analog, whereas the station beams are beamformed in the digital domain. Each element is beamformed and summed into two independent beams. Analog beamforming using true time delays is developed and demonstrated [RD-12].

Advantages:

- Cost and power are reduced due to a large reduction of required signal paths, which need to be digitized and processed (fewer receivers, anti-alias filters, ADCs and digital processing required)
- Received RFI is reduced by the reduction of FOV before the ADC
- Can be done close to the antennas

Disadvantages:

- Since this is analog, imperfections are created. This results in a tiny gain loss.
- Needs low date rate control in the field near the antennas.

### 9.4.2 Digital Beamforming

The function of the digital beamformer is to time delay antenna signals and sum them thereafter. This can be done by either a time delay or phase shift. Usually the latter is chosen since somewhere in the chain frequency channelisation needs to be created to obtain a required frequency resolution and it offers the possibility to trade-off bandwidth for FoV. Hence, a filterbank is required in front of the digital beamformer. The number of channels created depends on the station size and the tolerable error at the edges of each channel.

Advantages:

- The beamforming is deterministic and known for each signal path. No variations are present due to temperature variations for example.
- It offers flexibility and extendibility in a later phase, if the cost of processing keeps going down

Disadvantages:

- Each antenna element signal needs to be transported, digitized and filtered into channels prior to the digital beamformer. This results in processing power, extra costs and power.
- If the digitization is not done near the antennas, more calibration is needed for the signal paths which requires a significantly larger station correlator.

## 9.5 Distributed versus Centralised Architecture

The antennas of MFAA are spread over a large area and need to be processed. The location of the processing is a design parameter. The processing can be distributed with possibly several stages or in a central location.

### 9.5.1 Distributed Architecture

In LOFAR a distributed architecture was used, wherein a physical station matches with a logical station. The main function of the concept of stations in LOFAR was to reduce the amount of data as soon as possible in the signal chain to save on transport cost. For that each station has a number of cabinets installed wherein the beamformer is housed. From there on the station signals are sent to a central location, which is housing the central signal processor typically for correlation.

Advantages:

- Cost saving by reducing the data as soon as possible in the signal chain
- No need for an expensive large building to house all the electronics

Disadvantages:

- Maintainability of the equipment is distributed over stations
- Distribution of power, cooling and communications
- Station size is fixed (although there are solutions to circumvent this)

### 9.5.2 Centralised Architecture

For LFAA the core antennas are all connected to a central processing facility. This enables the antennas in the core to be made into large areas of antennas. Therefore, a logical station does not need to be the same as a physical station.

Advantages:

- Very flexible in the way stations are defined
- Electronics can be housed in a temperature and humidity conditioned environment
- Maintenance can be done more easily in one building
- Ability to implement advanced processing approaches

Disadvantages:

- Cost of data transport over a longer distance
- Installation more complex due to the large amount of cables going in the central building

## 9.6 Sparse/Irregular Versus Dense/Regular

Two configurations are being considered for MFAA. A configuration wherein the antennas are separated further apart: sparse/irregular, and a configuration wherein the antennas are packed closely together dense/regular. Formally the term sparse and dense is not correct, since it should relate to a certain frequency. An array can operate in the dense regime or sparse regime. Assuming a pitch in between the antennas of $\lambda_p$ corresponding with a frequency $f_p$ then a design is in the

- Sparse regime for $f > f_p$
- Dense regime for $f < f_p$

In this document the term sparse array is used when the sparse regime is larger than the dense regime for the frequency band of interest (450 – 1450 MHz).

### 9.6.1 Sparse/Irregular Array

In a sparse array the antennas are distributed further apart. By using a sparse array, signals with a frequency larger than the pitch frequency are under-sampled in the spatial domain causing grating lobes to appear. A way to reduce or scramble the grating lobes is by using an irregular array.
Advantages:

- High A/T for lower frequencies
- Fewer antennas required for the same performance at lower part of the band

Disadvantages:

- A/T drops with $f^{2}$ profile within the band of interest
- Beam model less smooth and predicable as within a dense array
- Smeared grating lobes within the band

### 9.6.2 Dense/Regular Array

In a dense array the antennas are closely packed. The pitch is chosen such that a first grating lobe enters the band at the top end of the band for the maximal scan angle. Due to the regularity of the array the mutual coupling in between the antennas is the same for all antennas, except for the boundary antennas. This makes the station beam pattern very stable, smooth and predictable. The station beam pattern is basically defined by the station geometry instead of the individual antennas.
Advantages:

- Station beam pattern very stable, smooth and predictable
- Far outside-lobes

Disadvantages:

- A curved A/T performance, which gives only nominal performance in the centre of the band
- A large number of antennas is required, since the pitch is driven by a higher frequency and the antennas are smaller

# 10 System Concepts

In this section two system concepts are presented which combine the physical design solutions discussed in Section 9 in a full system concept. Also the results from the costing document are presented.

## 10.1 System Concept 1

The first system concept is inspired from the EMBRACE [RD-5] demonstrator.  Figure 22 shows a system level overview of one station. Within this concept analog tiles are used, wherein an analog beamformer reduces the amount of signals significantly. The antenna array is a dense array with a pitch of 12.5 cm. The station processing is done in a hut next to the antenna array (distributed processing) in order to reduce the amount of data to transport further as soon as possible.

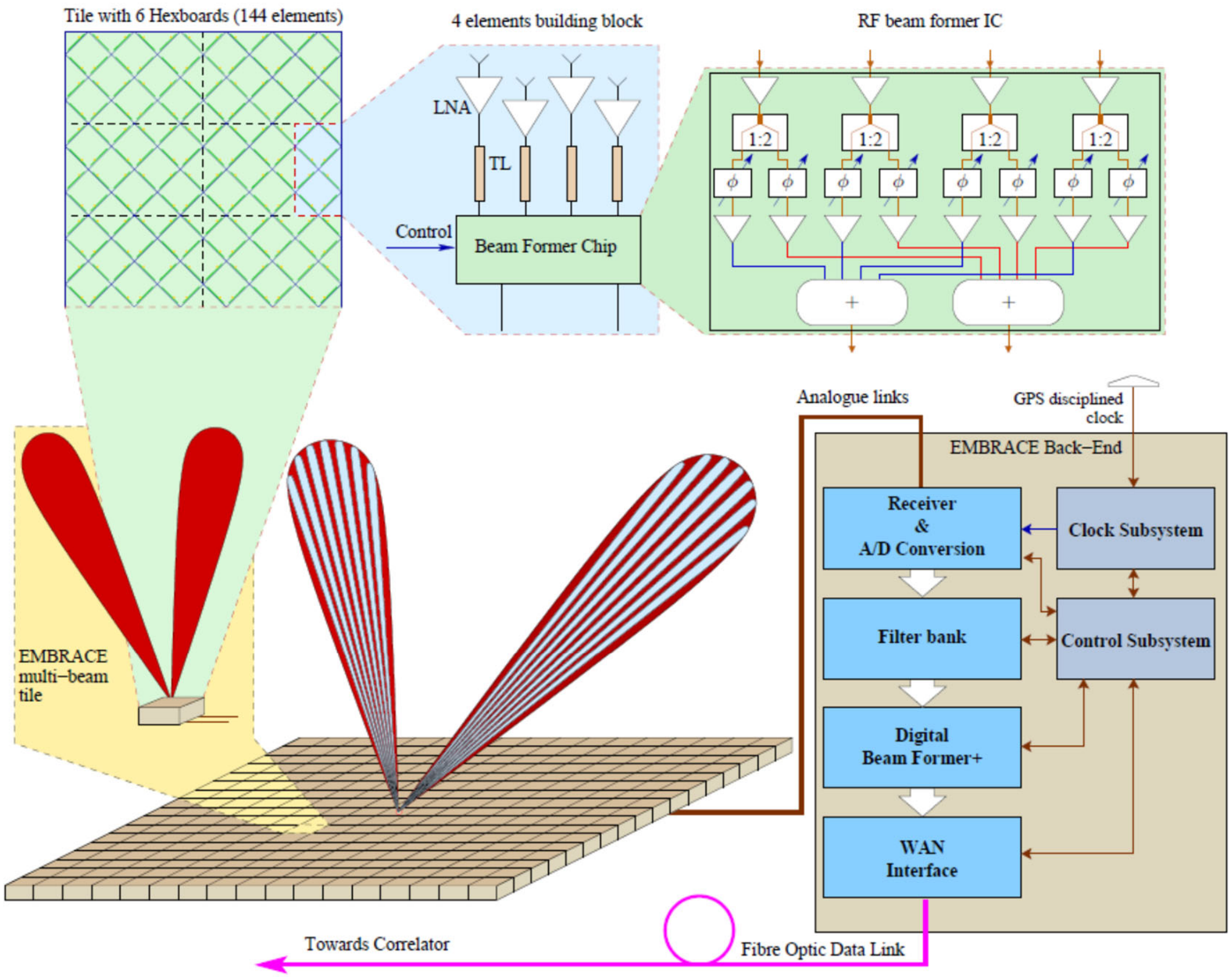

**Figure 22: EMBRACE concept**

A station can be divided roughly into two parts: a front-end and a back-end. The front-end consists of the antenna array including the radome, the supporting mechanics for the array, Low Noise Amplifier (LNA), analog beamforming, gain stages prior to signal transport, and the signal distribution between the tiles and the back-end. Since active LNAs have to be integrated close to the elements, the distribution of DC power is an important cost driver. The number of required cables can be reduced by multiplexing DC power, control signal and an analog RF beam signal on one coaxial cable. The analog beamforming can be done by phase shifters or delay lines. The signals of multiple antennas within a tile are typically combined in several stages. The back-end handles signal conditioning, analog-to-digital conversion and digital station beamforming. The station beams are transported to a facility for correlation either on the site or back to the post processing facility via a Wide Area Network (WAN).

## 10.2 System Concept 2

The second system concept is inspired by the LFAA architecture. It is a similar architecture as LFAA, but designed for a higher frequency range with obviously smaller antenna elements.

The basis of the core antenna placement and processing facility in this approach is shown in Figure 23, which is a possible overall MFAA core layout. The entire Core Area and more distant stations, within a radius of ~10-20km, are connected to a, single Central Processing Facility (CPF). Antenna stations on the spiral arms at greater distances than ~10km are connected to Remote Processing Facilities (RPFs). The RPFs consolidate stations within the range of the analogue communications or maybe the system only requires a *single* processing facility for the entire array. This is a conceptually simple arrangement that is highly flexible, upgradeable and easily maintained. The design could locate a correlator and clock source in the same CPF as the MFAA beamforming hardware. In this concept randomly placed individual log-periodic antennas are assumed. The antennas are optimised for the close antenna spacing required, dual polarisation performance and low cost. Each antenna system has the associated analogue electronics: low noise amplifier, gain with filtering.

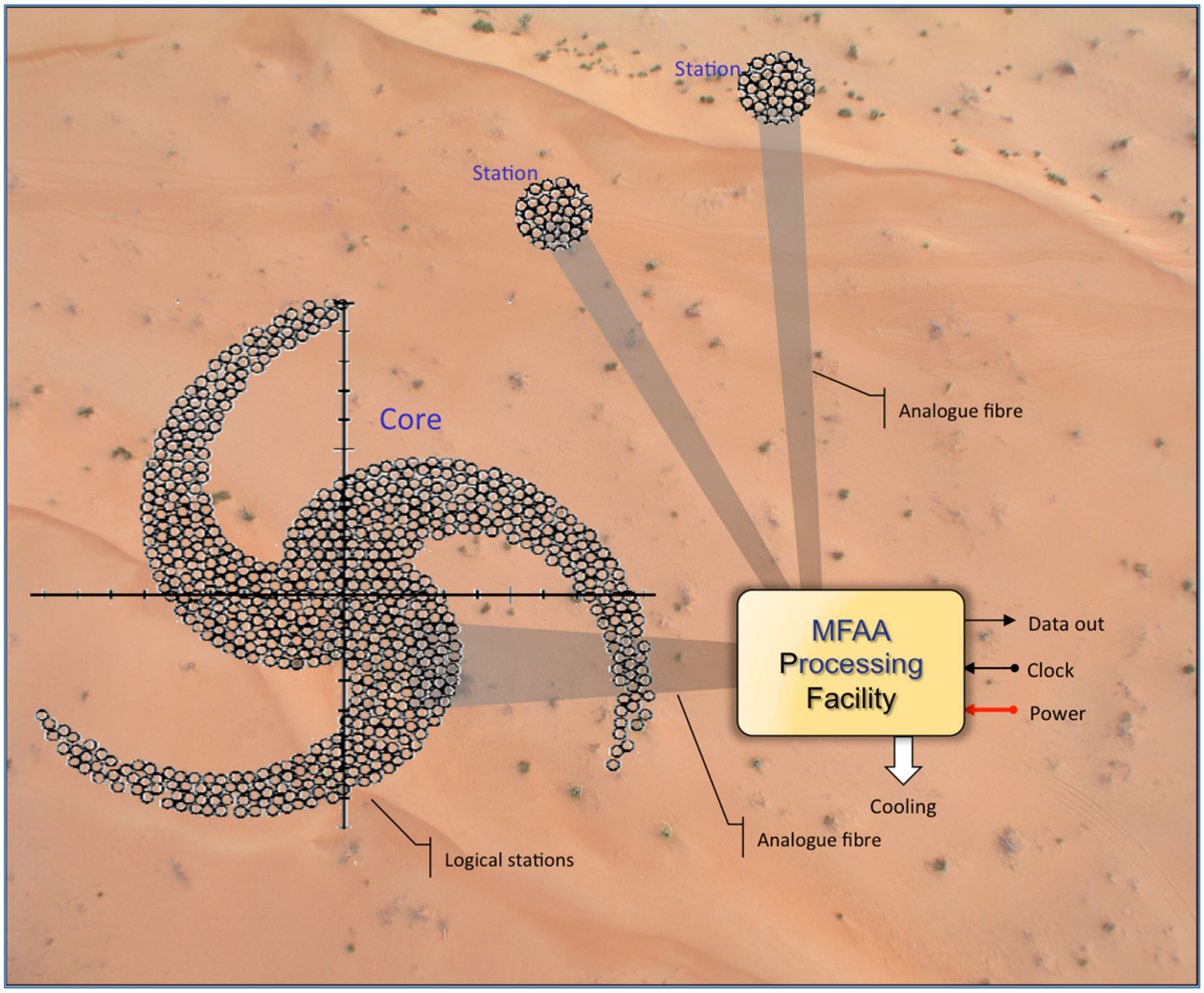

**Figure 23: MFAA core configuration proposal, with inner stations**

Within the processing facilities the analogue signals are digitised in frequency and calibrated; digital tile beams are formed using the required bandwidth. The processing "unit" in the CPF is the Tile Processing Module, TPM. All the TPMs are linked on the MFAA Data Network (MFAA-DN). The tile beams are combined into logical "station" beams of the selected station size and finally passed on for correlation.

The RPFs use essentially the same beamforming hardware as the CPF. Data flow between the RPFs and CPF uses digital fibre links implemented as an extension of the MFAA-DN.

The connection topology of the network interconnecting the TPMs and the control processors is shown in Figure 24, which shows a sparse array – this is configured as an all-digital system with signal communications *only* from the antennas to the processing facility.

As can be seen, all the data is carried over the MFAA-DN network including the control, monitoring and calibration data as well as signal data.

A cluster of high performance servers, which are connected to the network, controls the MFAA. These are the "Monitor, Control and Calibration System", MCCS, servers; they perform all the control functions for MFAA and link directly to Telescope Manager, TM, of the AA-MID telescope for instruction and reporting.

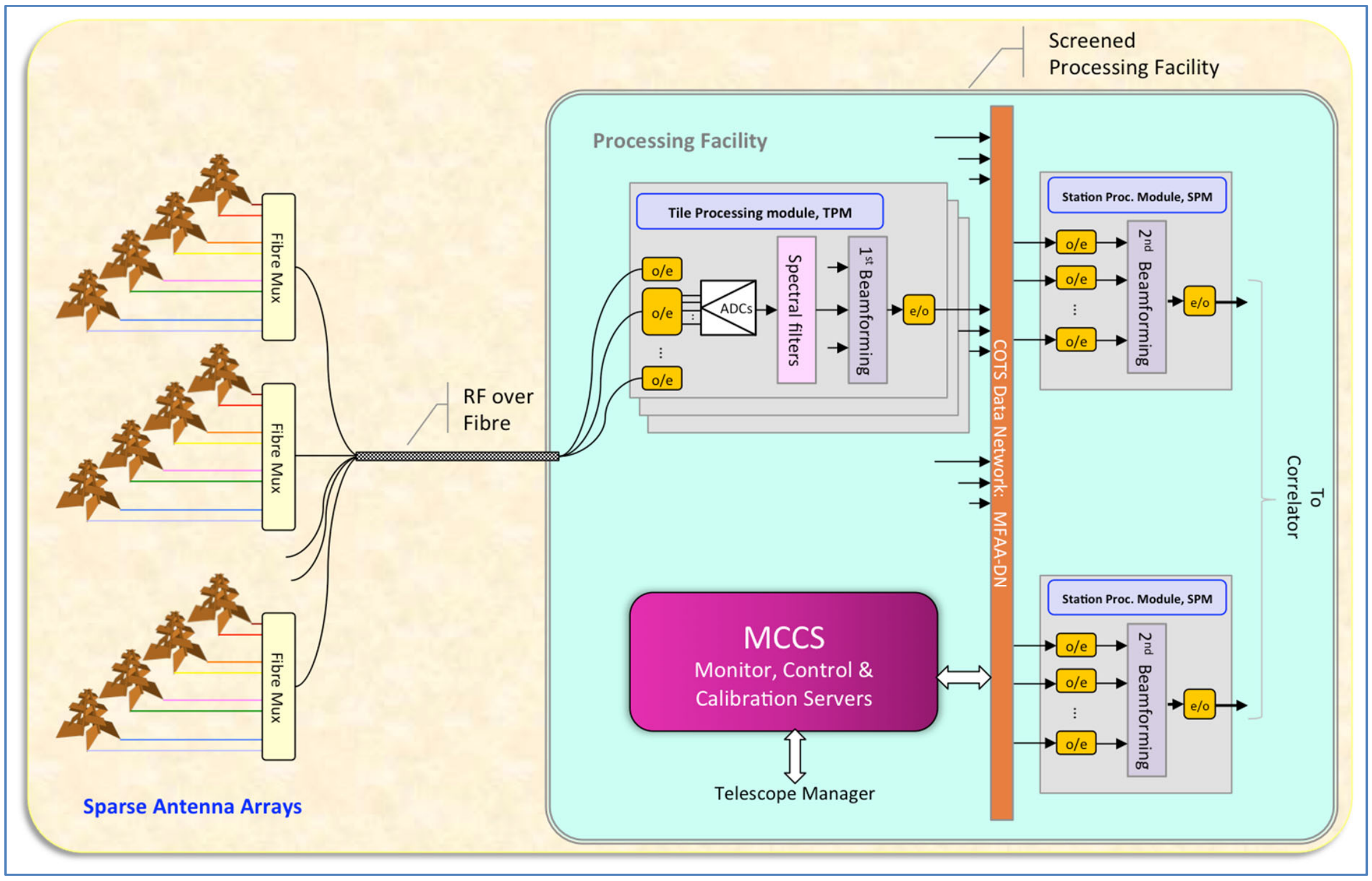

**Figure 24: Outline architecture using a sparse array**

The calibration of the analogue receiver chains, which is crucial for the high dynamic range and other performance of the AA-MID, is controlled and calculated within the MCCS. In essence, some amount of antenna data, from every antenna in every station is sent to the MCCS, which performs antenna level calculations to determine the correction coefficients for amplitude and relative phase to correct the pass-bands of every signal chain; these are communicated back over the MFAA-DN.

Links from the MFAA-DN or the station processors transfer the beam signal data to the post processing systems.

This system incorporates two levels of digital beamforming by using the TPMs to produce "Tile beams" which are passed to Station Processing Modules, SPMs. The SPMs can produce many station beams within the beam size of the tile. This is only necessary for very large numbers of beams and depends critically on the ability of the post processing system to handle this much data. Until the pot processing can accommodate this data rate then the system can be simplified by not having SPMs and using a "daisy chain" approach of accumulating beams on the TPMs. This has the benefit that every beam is "perfect" in that it has been formed by a distributed beamformer. This is can be used up to the bandwidth capabilities of the MFAA-DN. Unless, of course, more advanced processing techniques are developed.

## 10.3 Performance Comparison

Given the same cost [RD-13] the performance of the regular dense array (system concept 1) and irregular sparse array (system concept 2) are given in Figure 25 and Figure 26. The difference in performance is shown in Figure 27 at zenith and at 45 degrees scan angle. The performance of the regular dense array is lower for the lower frequencies, but higher for the higher frequencies. Figure 27 clearly shows the effect of scan angle on the sensitivity.

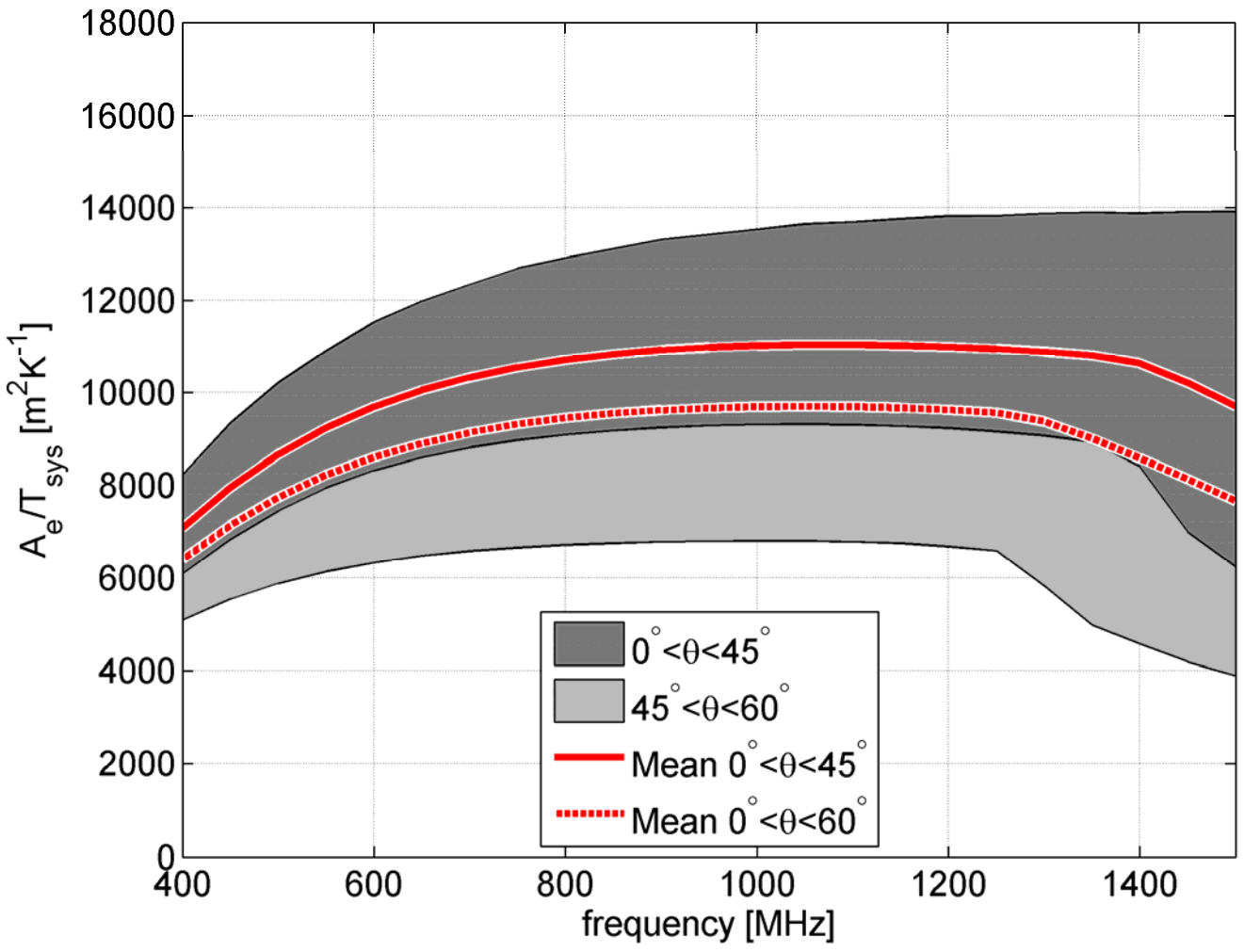

**Figure 25. Modelled sensitivity of MFAA implementation A (regular dense)**

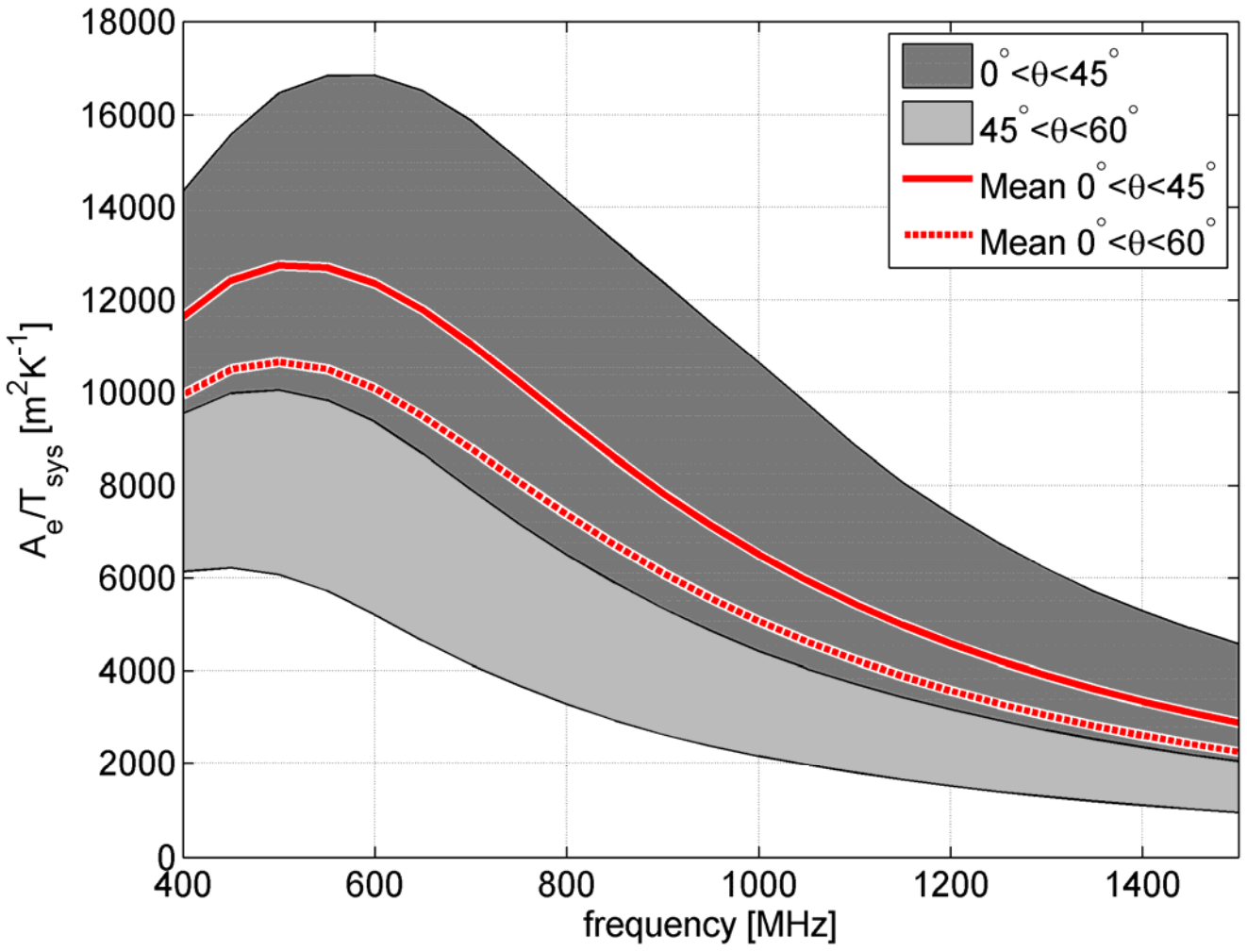

**Figure 26. Modelled sensitivity of MFAA implementation B (irregular sparse)**

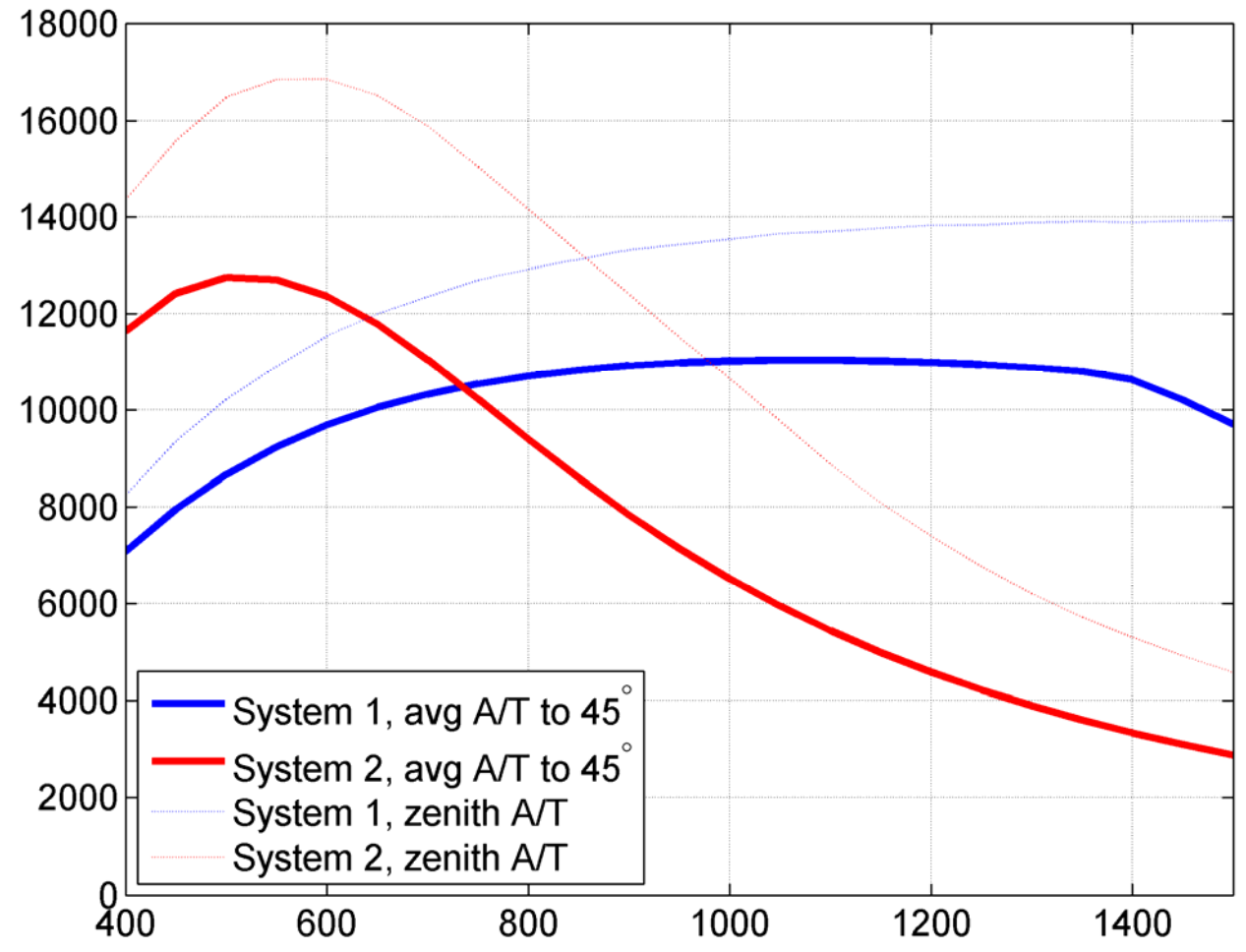

**Figure 27. Sensitivity comparison of two equal cost MFAA implementations**

# 11 Appendix Preliminary AAMID System Optimization

In this Appendix results are presented from a preliminary study covering the complete AAMID system to investigate its feasibility system wide. In the first section the computing requirements are discussed. This gives a first order estimate of the total computing required for the full system. For that a number of assumptions are used from existing work based on LFAA and the algorithms used there. In the following section an optimization of the number of stations is considered based on the required amount of processing, calibration and UV coverage requirements. The last section discusses an opportunity for AAMID to take advantage of dishes for the outer stations.

## 11.1 Computing Requirements

In this section the computing requirements are calculated for the three major processing steps of the instrument: MFAA processing, correlator processing and science data processing. For each of the steps, the computing requirements are determined in number of arithmetic operations per second required for real-time imaging of MFAA data.

### 11.1.1 Assumptions

The processing is one of the driving factors for the beamformer, correlator and science data processing. Bandwidth and memory costs are not assumed at this stage. Processing is assumed as metric since it has been one of the driving cost factors in existing systems and can be used to estimate the power of the AAMID system as well.

The type of technology used depends on the processing complexity required. Since datarates are significant at the front of phased array systems, usually fixed point processing engines as FPGAs are used there. Further down the signal chain, usually GPU or commercial computer systems are used which process data in floating point format. In this section the processing is expressed in number of arithmetic operations per second, no distinction is made between floating-point or integer operations. As a result, care should be taken when comparing processing in different processing stages: floating-point arithmetic consumes more power than integer arithmetic, but the cost of energy is likely higher in the desert.

For these calculations is assumed that we make continuum images using the full array (80 km baselines assumed) with 250 MHz bandwidth between 400 and 650 MHz. In the image cube, we make one image for each of the 262,144 channels. The rationale for this high frequency resolution of the images is that it is expected that a high resolution will be needed for calibration.

The computing model we use is based on work by [RD-10] and [RD-11]. The imaging algorithm is assumed to be w-snapshots.

By future optimising the design of MFAA significant savings in processing requirements for the overall AAMID telescope can be made.

### 11.1.2 MFAA Processing

For MFAA processing we model the computing cost of the polyphase filters, beamforming, station correlation for calibration, and calibration. The polyphase filters generate 512 channels each. For the FIR filter 16 taps is assumed. Calculating the full correlation matrix for calibration is performed every 4 minutes.

### 11.1.3 Correlator Processing

In the correlator, the computing cost of the polyphase filters, correlation, and an RFI flagging stage are modeled. The polyphase filters (16 taps per FIR) increases the frequency resolution to a total of 262,144 channels (512 channels per subband). The integration time for each of the designs is calculated as $1200*D/B_{max}$, with $D$ the station diameter and $B_{max}$ the maximum baseline.

### 11.1.4 Science Data Processing

The science data processor contains many algorithms. The imager contains three cycles: the calibration cycle, the major cycle and the minor cycle. The goal of each cycle is either to improve the sky model or to apply new calibration parameters based on the current sky model.

Visibilities received from the correlator are first calibrated using an initial sky model. Calibrated visibilities are then processed in the first gridder and an inverse Fourier transform generates the first (dirty) image cube. A number of minor cycles are executed on this image cube to extract sources in the image domain and improve the sky model. However, source subtraction in the image domain is not perfect and, as a result, the major cycle needs to be finished by going back to the original visibilities and subtract all sources found in the minor cycles in the u,v-domain. This is gridded and Fourier transformed again to generate a new dirty image cube.

The process of several minor and major cycles repeats until sufficient sources are subtracted to finish a calibration cycle. Based on the new sky model, new calibration parameters are generated and applied to the visibilities. With the newly calibrated visibilities, new major and minor cycles are started.

Although the exact calibration strategy for AAMID is not yet known, a first order estimate is found by incorporating the model for direction-independent and direction-dependent calibration derived for SKA1-Low. Besides the computing cost for calibration, the computing cost for gridding, A-projection, Fourier transform, re-projection for w-snapshots, and PSF source subtract is included.

For the analysis, three calibration cycles are set with each 10 major cycles and 100 minor cycles per major cycle. For calibration, 300 sources and 30 directions are set. The number of pixels in the image is calculated as $3*B_{max}/D$. The snapshot time for w-snapshots is optimized for minimum computing requirements. A-Projection kernels are recomputed every 5 minutes and frequency stability is set to 4 MHz. Furthermore, immediately before gridding a two-zone baseline dependent averaging is applied. For baselines smaller than 6 km, first integration over time is done before gridding as this reduces the computational rate.

### 11.1.5 AAMID Computing Requirements

#### *11.1.5.1 A/T versus FoV Trade-off*

The AAMID computing requirements for the designs presented in Section 7.3.2 are depicted in Figure 28. The design which is optimized for a large A/T uses significantly less processing than the design which is optimized for FoV. The design which is optimized for FoV uses many more beams, which need to be formed in the beamformer, correlated in the Central Signal Processor (CSP) and further processed in the Science Data Processor (SDP). All those processing steps scale linearly with the number of beams. The processing is about an order of magnitude larger which is due to the number of beams which is also an order of magntitude larger compared with the A/T design. The main contribution in the phased-array processing (station processing) is the beamforming.

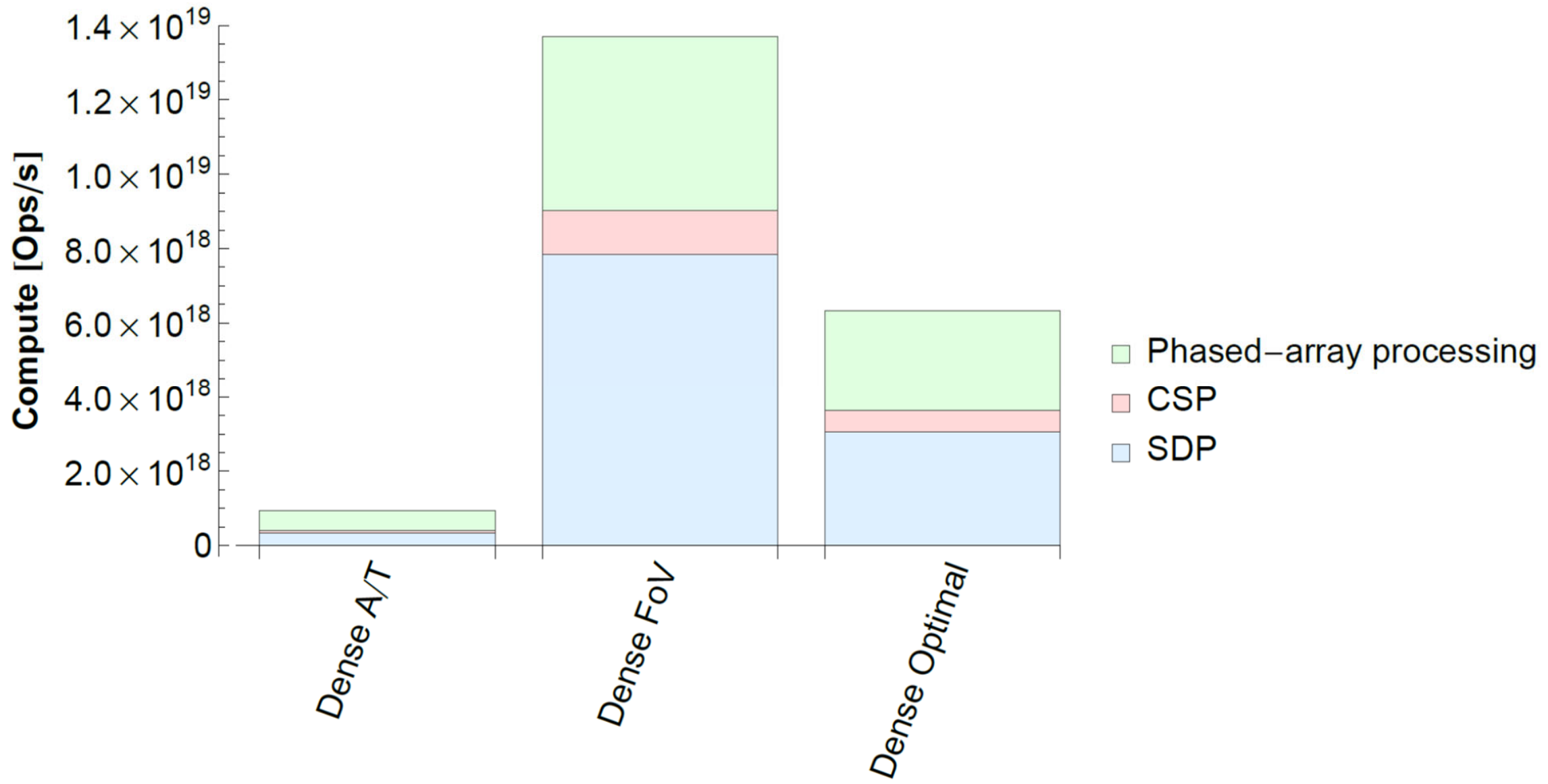

**Figure 28: Compute requirements for the dense designs wherein respectively the A/T is maximized, the FoV is maximized and an optimal design**

Figure 29 shows the distribution of processing load in the science data processor. Clearly the direction-dependent calibration is the most dominant processing step. It should be noted again, that a number of assumptions are used to produce these results. Furthermore, there will be most likely important developments to come, which need to be incorporated in this model.

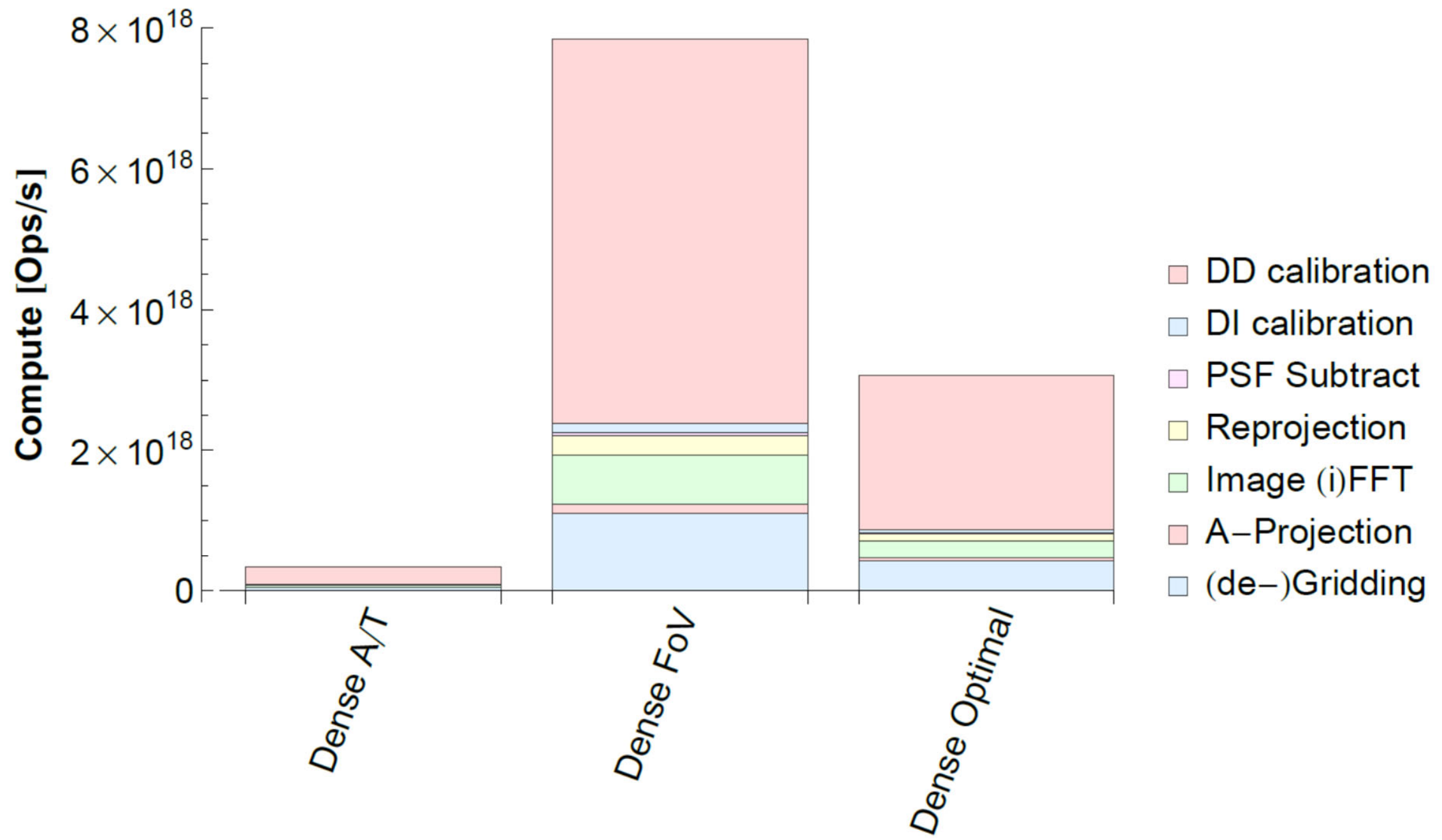

**Figure 29: Science data processor compute requirements for the dense designs wherein respectively the A/T is maximized, the FoV is maximized and an optimal design**

### *11.1.5.2 Dense-regular versus Sparse-irregular Trade-off*

The AAMID computing requirements for an all-digital design and dense "Optimal" design is presented Figure 30. In the all-digital design all beamforming is done in the digital domain. Two stages of beamforming are assumed, wherein 12 antennas are combined in the first stage.

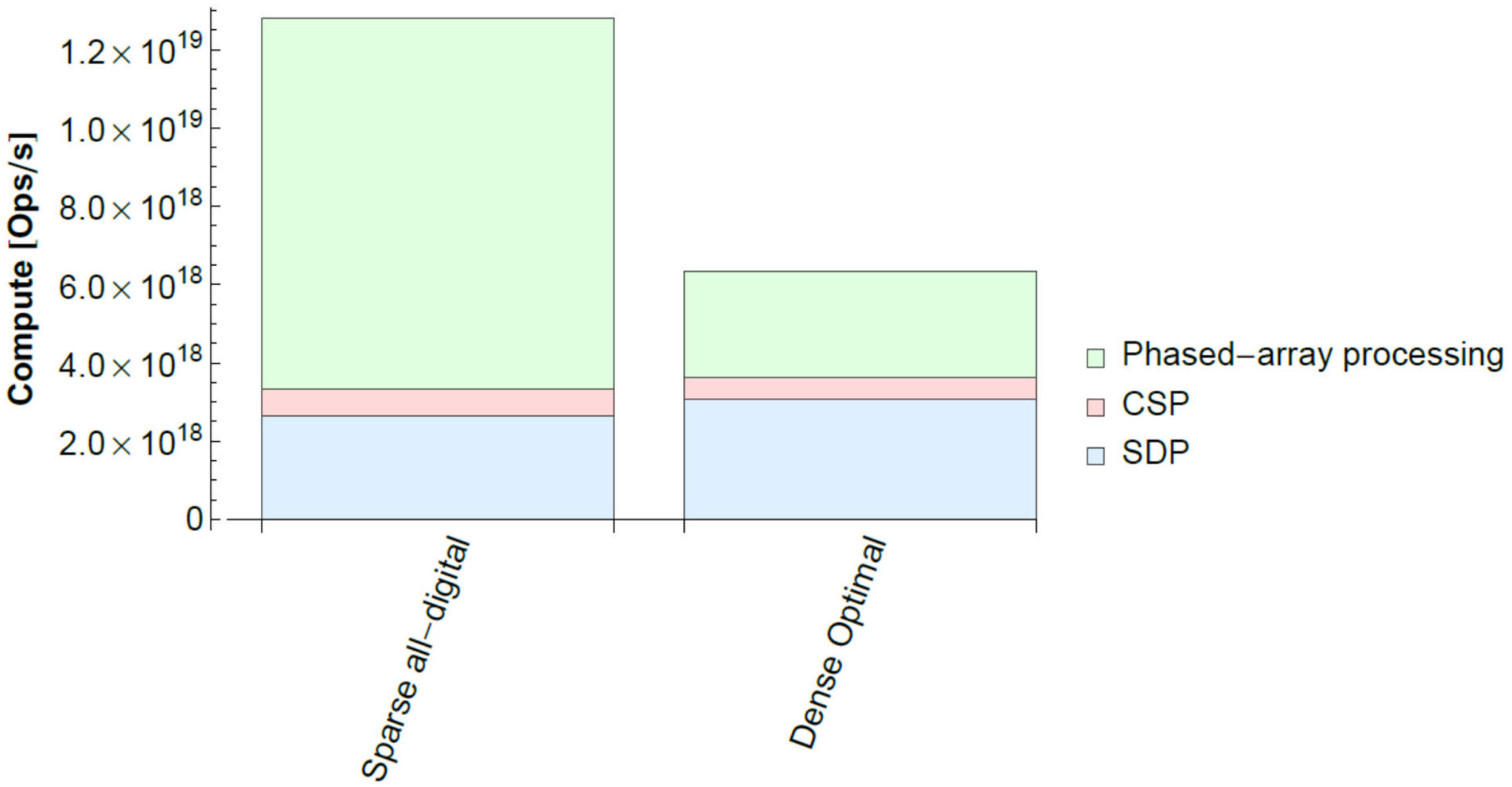

**Figure 30: Compute requirements for a all-digital sparse design and the dense "Optimal" design**

Although the sparse design has much less antenna elements, still the beamformer processing takes twice the number of compute resources compared with the dense design. This is caused by the fact that the number of antennas which can be combined to obtain the optical FoV for the sparse design is much less (12) than for the dense design (72). Therefore, the number of inputs to the digital beamformer is larger. Furthermore, the station area is larger for the sparse design which causes smaller station beams. To gain the same processed FoV more digital beams needs to be created. This same argument results in less science data processing for sparse, since the most demanding functions go with $D^{-2}$, wherein *D* is the station diameter.

The distribution of phased-array processing for these designs within a station is depicted in Figure 31. Interesting to note is that the digital beamformer is not even a factor of two higher than for the analog beamformer. Still the correlation load is severe under the used assumptions, but this can probably be reduced by using alternative schemes for the calibration, which need further investigations. The model used assumed that each individual signal path needs to be calibrated because each signal needs to be transported over quite some distance to a central location for a station or perhaps a central location for the whole core.

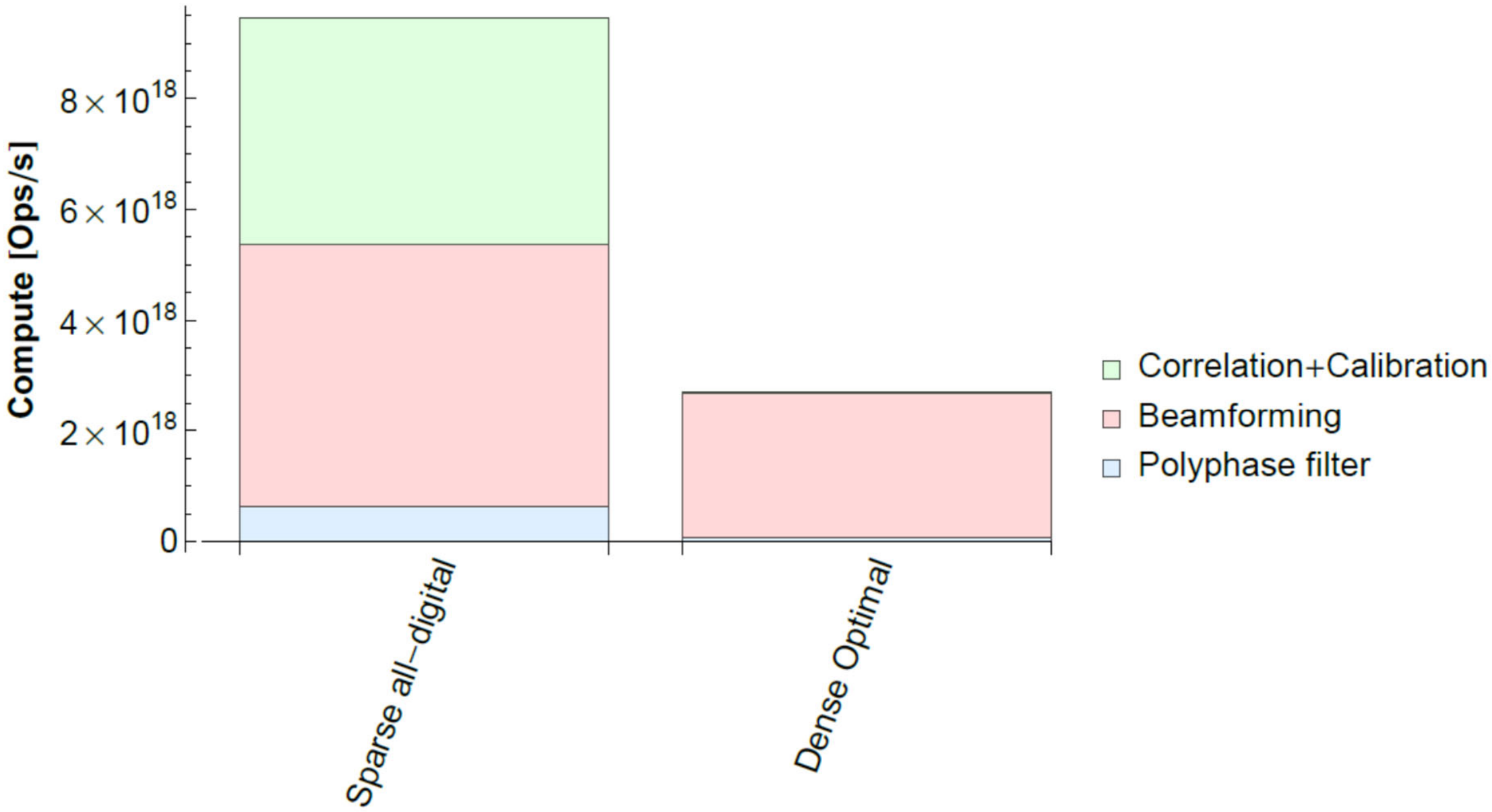

**Figure 31: Phased-array processing compute for a sparse design, all-digital sparse design and the dense "Optimal" design**

## 11.2 Optimizing the Number of Stations

Given a design with a fixed FoV and sensitivity still leaves the number of stations as a "free" parameter. The required number of stations is driven by the following parameters:

1. Required instantaneous UV coverage
2. Required synthesis UV coverage
3. Minimum baseline required
4. Minimal station size required for ionospheric calibration
5. Minimal station size required for gain/phase calibration of individual antennas
6. Processing costs of the beamformer, correlator and science data processing

These topics are further addressed in the following subsections.

### 11.2.1 Instantaneous UV Coverage

The required instantaneous UV coverage is driven by the following requirement:
AM_R3043 stating that AAMID shall have sufficient instantaneous UV coverage to allow localization of fast transients in imaging data on imaging time scales of no more than tens of milliseconds and as short as 1 ms.

### 11.2.2 Synthesis UV Coverage

Currently there is no specific requirement targeting a required UV coverage from a synthesis imaging requirement. This needs to be further specified and detailed in the next phase.

### 11.2.3 Minimum Baseline

Currently the minimum baseline required is 20 m and preferably even 1 m (which can only be conceived of in an aperture array). This is probably only required for the core, but can still lead to a significant number of stations. More detailed study is necessary to determine the number of stations required for these science cases.

### 11.2.4 Ionospheric Calibration

Based on an analysis presented in RD-3, the number of sources in the main beam area with a useful SNR (assumed to be 5) for calibration as well as the number of sources required to characterize the ionospheric phase screen over each individual station can be calculated. For this analysis, the total installed collecting area is assumed to be constant and equivalent to the collecting area of 250 stations with 53-m diameter. This means that smaller stations will lead to more stations to retain the same total collecting area. Furthermore, the following assumptions were made

- Calibration is performed per 1 MHz bandwidth with 10 s update rate;
- The receiver temperature if 40 K;
- A typical traveling ionospheric disturbance has a wavelength of 120 km and an amplitude of 0.1 TECU;
- The ionospheric phase screen is placed at 200 km height above the array.

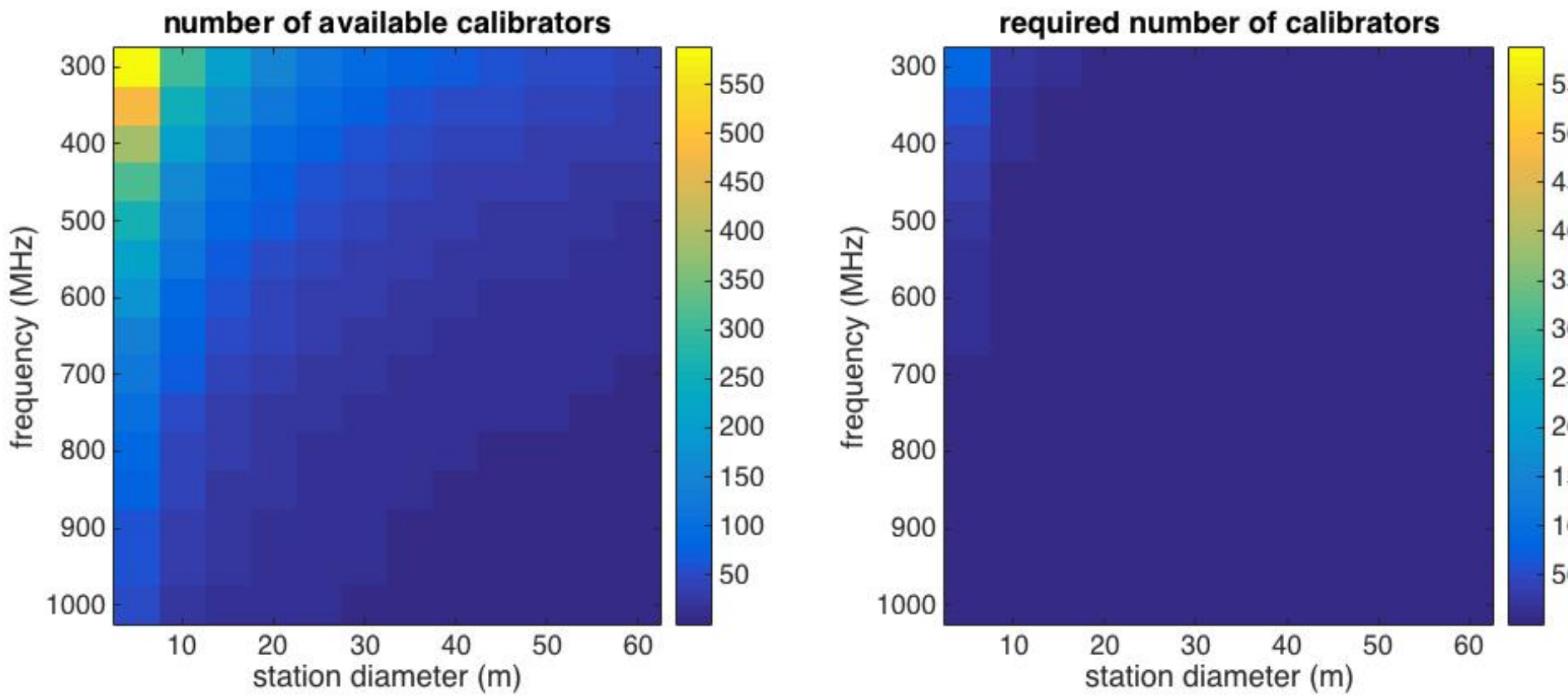

**Figure 32: Number of available calibration sources and number of required calibration sources for ionospheric characterization as function of frequency and station diameter**

Figure 32 shows the number of available calibration sources, i.e., sources with SNR larger than 5, as well as the number of calibration sources required for calibration as a function of frequency and station diameter. The two images are plotted using the same color scale. By comparing these two images, we can easily see that for low frequencies and small station sizes, the number of available calibration sources is higher than the required number of calibration sources. We may therefore conclude that ionospheric calibration poses no lower limit on station size when the envisaged total collecting area is installed. Even for observations with an array consisting of 5-m stations at 300 MHz, the number of available sources is more than a factor 5 higher than the required number of sources. This implies that even if the collecting area is reduced by a factor 2 or 3, the conclusion that ionospheric calibration does not put a lower limit on station size, still holds. For a system of the envisaged size, the trade-off between number of stations and station size can thus be made based on considerations such as instantaneous (u, v)-coverage and overall system cost without considering feasibility of characterizing the ionosphere unless the total collecting area is significantly reduced.

### 11.2.5 Gain and Phase Calibration

The signals from a limited number of antennas may be beamformed locally to form a tile signal, or sent from individual antenna elements. This can be implemented with an analog beamformer or a digital beamformer. In the latter case, from a calibration perspective, it is preferred that the ADCs are placed in the tiles and that care is taken about the clock distribution to these ADCs (see Section 9.1), however, that may be impractical due to excess interference being generated. This ensures that all links whose dielectric length is crucial for proper delay compensation in the tile have a length that is constrained to the tile size. This constraint ensures that variations in dielectric length due to temperature variations are negligible, which implies that, in principle, a tile only needs to be characterized once over its lifetime. The physical length (and therefore also the dielectrical length) of the links between the tiles in the field to a station or central signal processing unit suffer from significant variations in dielectric lengths due to temperature variations, that need to be corrected by calibration. If, in case of the digital beamformer, all antenna signals are sent to a central location extra risk is involved in calibrating for the dielectrical length variations.

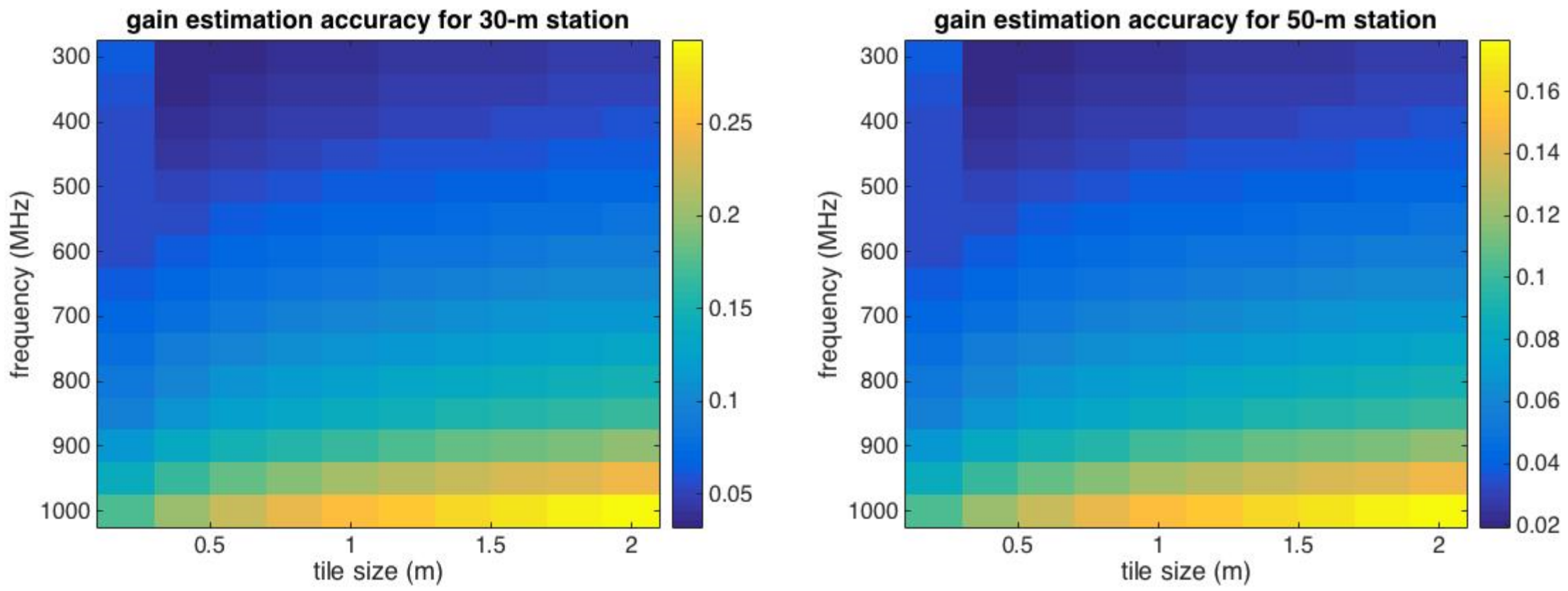

**Figure 33: Receiver path gain calibration accuracy (as fractions) as function of frequency and tile size for a 30-m and a 50-m station.**

If the station size is kept constant, the number of tiles in a station will increase if the tiles are made smaller. A smaller tile provides a larger FoV, which improves the odds of a strong source being present somewhere in the tile beam, but it also has a smaller collecting area, which reduces the sensitivity per tile and per tile-tile baseline. Fortunately, the estimation accuracy for the receiver path gains depends on the SNR per tile-tile baseline and the number of baselines available per tile. Figure 33 shows the expected receiver path gain estimation accuracy as function of tile size and frequency assuming calibration on the brightest source in the FoV, which is determined based on source statistic, for a 30-m and a 50-m station. In these simulations a bandwidth of 1 MHz and an integration time of 10 s was assumed. Using a larger bandwidth or longer integration time may improve the accuracy of the gain estimates further as may exploitation of the combined flux of other, less bright, sources in the FoV. These results therefore indicate that for AAMID stations of reasonable size (30-m diameter and larger), it is feasible to monitor the receiver path gains during an observation without pausing the observation to observe a bright calibrator.

### 11.2.6 Processings Costs

The relations used for calculating the computational requirements in Section 11.1 can be used as well to optimize the number of stations in the system to minimize the total amount of processing. For this analysis, the total installed collecting area is assumed to be constant and equivalent to the collecting area of the dense "Optimal" design. Therefore, smaller stations will lead to more stations to retain the same total collecting area. Also the FoV is assumed to be constant. This means that smaller stations need less beams to cover the same FoV compared with large stations. By assuming a constant A/T and FoV of the system, all processing relations can be re-formulated to show the dependency on the number of stations $N_s$. By doing so, the following dependencies are found (processing term is indicated in brackets):

MFAA

- filterbank: constant (Pfbstat)
- beamformer: $N_s^{-1}$ (Pb)

Central signal processor

- filterbank: constant (Pcorrfb)
- correlator: $N_s$ (Pcorr)

Science data processor

- gridder: $N_s^2$ (Pgrid)
- iFFT: $N_s$ (Pifft)
- direction independent calibration: $N_s^2$ and $N_s$ (Pdi)
- direction dependent calibration: $N_s^2$ and $N_s$ (Pdd)

For a maximal baseline length of 80 km, the processing curves are depicted in Figure 34. For the MFAA and central signal processor processing a FPGA implementation is assumed, wherein a very moderate 1 MAC/s = 2 Flops is taken. The total amount of processing is indicated by Ptot which is the black dashed line. From the figure can be concluded that the beamformer processing drives the optimal number of stations to a high number, while both the direction dependent and gridder drives the optimal number of stations to a lower number. The location of the optimum is close to 250 stations.

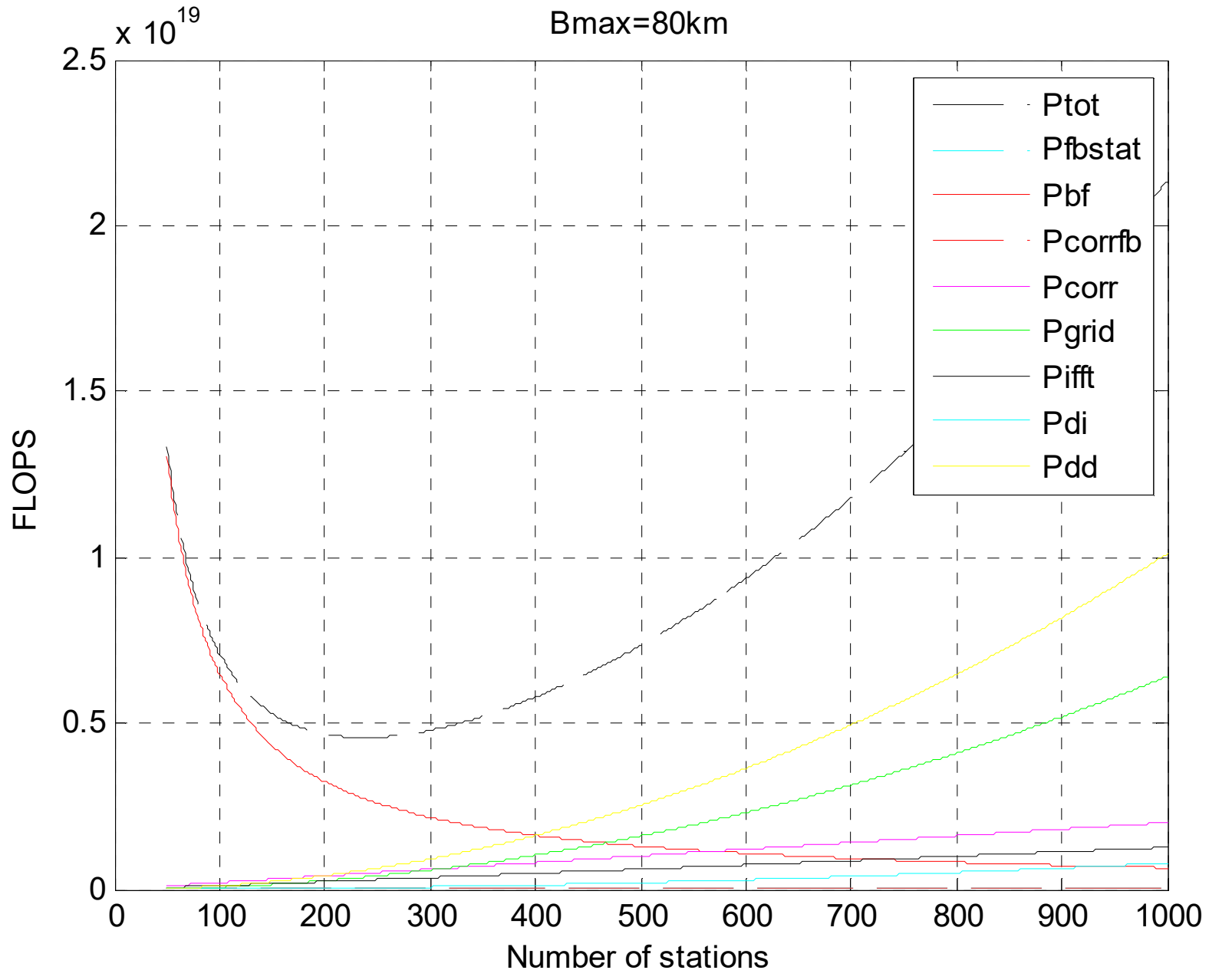

**Figure 34: Processing cost as function of the number of stations for a maximal baseline length of 80 km**

The maximal baseline length is a very crucial parameter for the science data processing load, since the gridder, direction independent and direction dependent calibration is growing with $B_{max}^2$, while the iFFT goes even with $B_{max}^4$. So, for larger baseline lengths also the iFFT starts to play a major role. Already for doubling the maximal baseline length to 160 km, this becomes obvious as is illustrated in

Figure 35. The result is that the optimal number of stations reduces by a factor of 2, to about 125 stations.

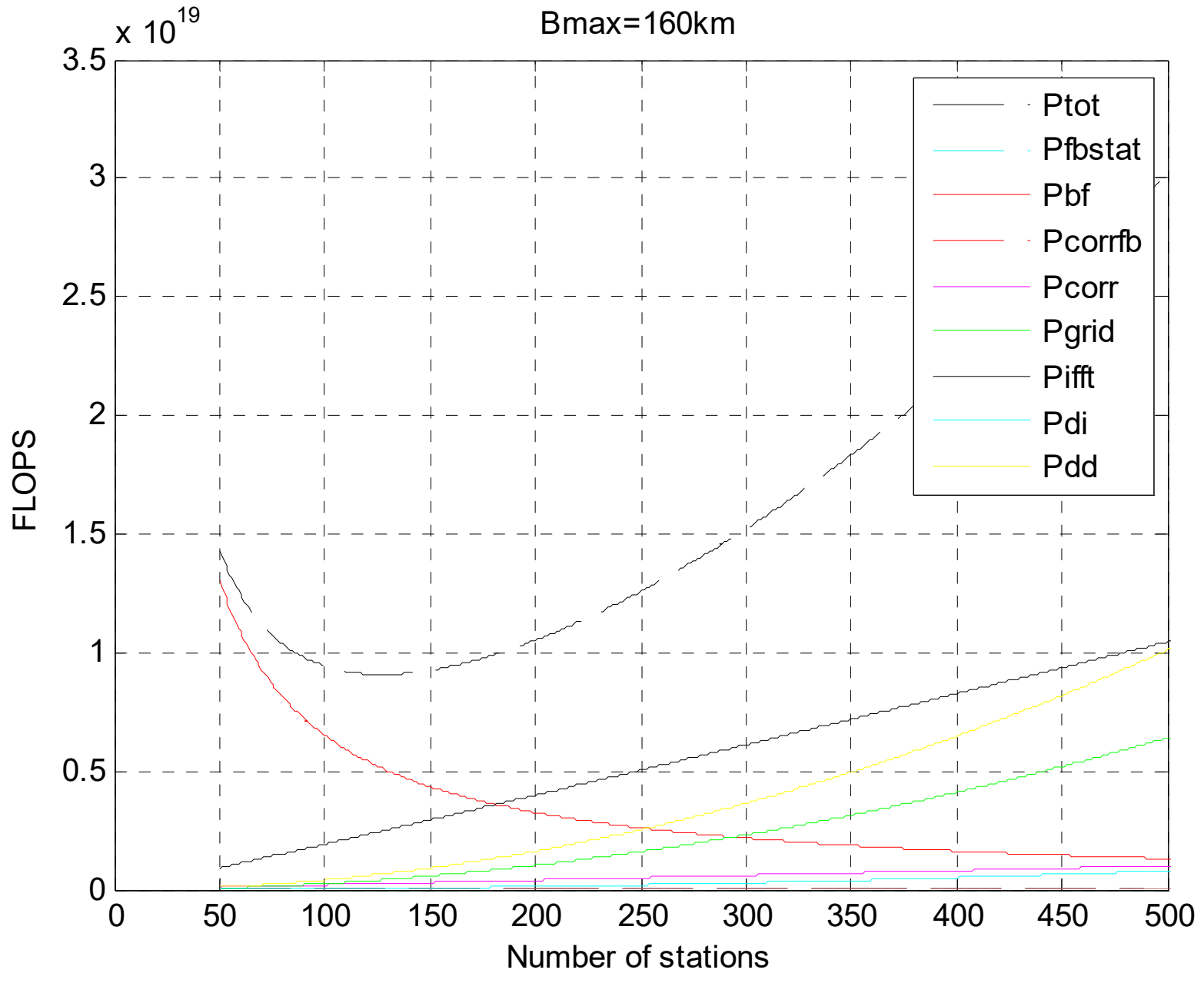

**Figure 35: Processing cost as function of the number of stations for a maximal baseline length of 160 km**

Since the requirement for the maximal baseline is currently 225 km, the optimal number of stations from a processing cost point of view is even lower as is shown in Figure 36.

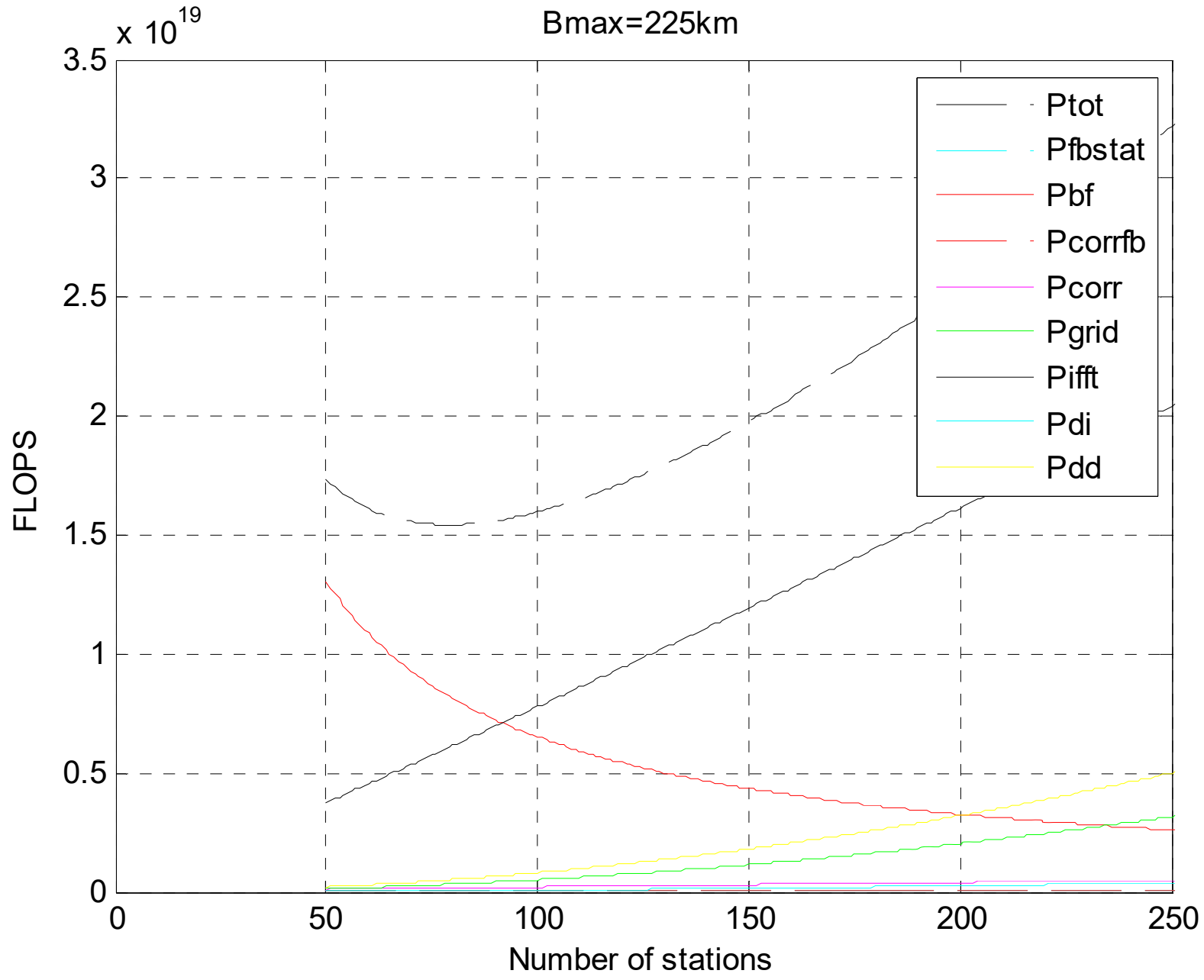

Figure 36: Processing cost as function of the number of stations for a maximal baseline length of 225 km

## 11.3 Mid Frequency with Aperture Array core stations and Dish outer stations

Several key science projects are optimally addressed with a large fraction of the collecting area in the core. These include pulsar searches, transient searches, the galactic HI survey, and the cosmology survey, especially the intensity mapping experiment for the detection of the baryonic acoustic oscillation signature. The Mid Frequency Aperture Array is perfectly suited to these surveys, providing an extremely large Field of View, and the possibility to correlate very short baselines for sensitivity to extended emission.

The large Field of View necessitates a significant digital processing capability, especially for surveys at high angular resolution. The combination of large field of view and high angular resolution results in a very large number of beams on the sky. In order to remain within processing limitations, only a subset of the field of view can be processed with high angular resolution. As a result, much of the RF capability of MFAA is unused when doing surveys with long baselines.

It is natural therefore to have an SKA architecture which is composed of MFAA stations in the core, and with dishes in the outer stations. Since the field of view is limited by processing capability, the limited field of view of the outer station dishes is no longer the constraining factor.

A hybrid architecture uses the advantages of both frontend technologies. The core MFAA provides very large field of view, while the outer stations increase the angular resolution for mapping within a subset of the MFAA field of view. Dishes are not limited to 45-degree scan angle, so the outer stations can follow the core field of view, even for stations at very long baselines.

A further consideration is the exploitation of existing instruments. When MFAA is being installed, there will already be MeerKAT and perhaps SKA1 operating. In the frequency overlap region, the new MFAA station can begin operations immediately as part of SKA1, providing a large core to the existing infrastructure. This will significantly augment the capability of SKA1 by providing increased sensitivity to the overall instrument, and, by beginning immediately a transient survey. SKA1 with MFAA will be a powerful instrument for transient detections and immediate follow-up observations.

In the nearer term, a MFAA technology demonstrator can be installed and running together with MeerKAT and KAT7. The combination would already be the most powerful transient survey instrument in operation, until SKA1 is operational.